\documentclass[10pt]{iopart}

\usepackage[ngerman,english]{babel}
\usepackage{graphicx}
\usepackage{subcaption}

\expandafter\let\csname equation*\endcsname\relax
\expandafter\let\csname endequation*\endcsname\relax
\usepackage{amsmath,amssymb}
\usepackage{bm}
\usepackage{siunitx}
\usepackage{isotope}
\usepackage{mhchem}

\usepackage{letltxmacro}
\makeatletter
\let\oldr@@t\r@@t
\def\r@@t#1#2{%
  \setbox0=\hbox{$\oldr@@t#1{#2\,}$}\dimen0=\ht0
  \advance\dimen0-0.2\ht0
  \setbox2=\hbox{\vrule height\ht0 depth -\dimen0}%
  {\box0\lower0.4pt\box2}}
\LetLtxMacro{\oldsqrt}{\sqrt}
\renewcommand*{\sqrt}[2][\ ]{\oldsqrt[#1]{#2}}
\makeatother

\usepackage{mathtools}

\newcommand{\IPP}{[1] Max-Planck-Institut f\"ur Plasmaphysik, Boltzmannstr.~2, 85748 Garching, Germany}

\newcommand{\ITER}{[2] ITER Organization, Route de Vinon-sur-Verdon, CS 90 046, 13067 Saint Paul-lez-Durance Cedex, France}

\newcount\colveccount
\newcommand*\colvec[1]{
        \global\colveccount#1
        \begin{pmatrix}
        \colvecnext
}
\def\colvecnext#1{
        #1
        \global\advance\colveccount-1
        \ifnum\colveccount>0
                \\
                \expandafter\colvecnext
        \else
                \end{pmatrix}
        \fi
}

\newcommand{\Alfv}{Alfv\'en}

\newcommand{\eg}{e.g.\ }
\newcommand{\ie}{i.e.\ }

\graphicspath{{images/}}

\begin{document}
\title{EP-Stability-WF: an IMAS-integrated workflow for energetic particle stability}
\author{V.-A.~Popa\textsuperscript{1}, Ph.~Lauber\textsuperscript{1}, T.~Hayward-Schneider\textsuperscript{1}, M.~Schneider\textsuperscript{2}, O.~Hoenen\textsuperscript{2} and S.~Pinches\textsuperscript{2}}
\address{ \IPP\\ \ITER}
\ead{alin.popa@ipp.mpg.de}
  \begin{abstract}
    The confinement of energetic particles (EPs) generated by fusion reactions and external heating methods is crucial for the performance of future fusion devices. However, EP transport can occur due to their interaction with electromagnetic perturbations, affecting heating efficiency and overall performance. Robust reduced models are needed to analyze stability and transport, but their development requires effort.

This paper presents an automated IMAS-based workflow for analyzing the time-dependent stability of EP-driven modes, focusing on the linear properties of Toroidal \Alfv{} Eigenmodes (TAEs) in general Tokamak geometry. The workflow utilizes efficient computational methods and reduced models to deliver fast and reproducible results. A demonstration of the workflow's effectiveness was performed, identifying key linear properties of TAEs in various projected ITER scenarios. This approach represents a critical step towards developing tools for analyzing EP transport and optimizing the performance of future fusion reactors.

  \end{abstract}
  \maketitle
  \ioptwocol 

  \section{Introduction}
Energetic Particles (EPs)  generated by fusion reactions and by  external heating methods such as Neutral Beam Injection (NBI)  need to be well confined in future (ITER, DEMO) fusion devices. However,the EPs' resonant and non-resonant  interaction with electromagnetic perturbations  can lead to EP transport which affects the heating efficiency and thus the overall performance of the burning plasma which is of concern  for future reactors.
Previous analyses show that in order to be able to determine the stability limits of weakly damped \Alfv{ic} perturbations and to model the related transport, a non-linear, global, kinetic treatment is necessary \cite{Lauber2013,ChenRMP2016}. But in order to study a whole scenario including ramp-up, flat-top and ramp-down that \eg{} at ITER is forseen to last $> 500$ seconds, one needs to  find a compromise between computational fidelity and speed.
Comprehensive codes with high fidelity (\eg{} ORB5, - global electromagnetic gyrokinetic particle-in-cell code~\cite{Lanti2020, Mishchenko2019} , GTC \cite{Lin1998} or GENE~\cite{GORLER2011}  can be used only for relatively short timescales (several ms), due to their extensive computational cost. Non-linear (gyro-) fluid models~\cite{Spong} can cover longer time scales, but besides the lack of kinetic effects the deep non-linear evolution poses still some unresolved challenges. Also direct coupling schemes of high-fidelity simulations and transport codes are presently developed~\cite{Di_Siena_2022}, that however are also extremely costly.
Therefore, robust  reduced models for the analysis of  stability and transport are needed as an additional tool. Presently, the following implementations based on different physics models are available: the critical gradient model \cite{Waltz_2015}, the  kick model \cite{Podest__2017},  the 'resonance broadening quasi-linear model' (RBQ) \cite{Gorelenkov_2018}, and the phase space zonal structure model (PSZS) \cite{Falessi_2019}.
While they are designed to be faster than comprehensive codes, still considerable effort is needed to determine the essential ingredients (at least linear mode features such as frequency, damping/growth rate, mode structure and related saturation rules) for the reduced models. In addition, to perform multiple linear computations at different time slices of a given scenario (ramp-up, flat-top and ramp-down) a high degree of automatization is required.

In this work, the first automated, time-dependent IMAS-based  (Integrated Modelling \& Analysis Suite (IMAS)~\cite{Imbeaux2015})  workflow (EP-Stability-WF) is presented and used to provide  insight into the stability of Toroidal \Alfv{} Eigenmodes(TAEs) in different phases of various scenarios. It is based on the equilibrium code HELENA ~\cite{HELENA} and different hierarchical models of the linear gyrokinetic eigenvalue code LIGKA\cite{Lauber2007} that are connected via a centralized IMAS database. A complex workflow (EP-Stability-WF~\cite{Popa_MSc}) was created to connect the different stages of the analysis with the database to ensure the reproducibility and consistent interpretation of the different analysis steps. With this tool, both time-dependent predictive scenario simulations and experimental data can be analyzed.

To demonstrate the capabilities of this workflow, several ITER scenarios for Deuterium Tritium plasmas are analyzed: a time-independent scenario generated by ASTRA~\cite{ASTRA} transport code, a time-dependant predictive scenario generated by METIS~\cite{Artaud_2018,Schneider_2021} transport code, and finally the newest ITER baseline scenario with 5.30 [T], 15 [MA] Deuterium Tritium plasma given by the DINA~\cite{DINA} - JINTRAC~\cite{JINTRAC} transport workflow which combines a free boundary equilibrium evolution code with the 1.5D core/2D SOL code. Here time-dependent means that an automated linear stability analysis is performed for different time points of an projected ITER scenario. These time points are sufficiently separated in time (typically 0.5s - 5s), i.e. it is assumed that the equilibrium evolution time is much slower than the \Alfv{} time scales connected to the linear and non-linear TAE physics.

The paper is organized as follows:  section \S\ref{sec:model} contains a technical description of the models and numerical tools, section \S\ref{sec:scenario}) gives an overview of the scenarios and parameters chosen for the computation. The presentation of the results (\S\ref{sec:results}) is split in three subsections: the evolution of damped TAEs without EPs, what happens to them when EPs are introduced, and a convergence test for a whole scenario. In the end, section \S\ref{sec:conclusions} presents the conclusions and outlook of the paper.

  \section{Numerical Tools and Physical model}
\label{sec:model}

\begin{figure}[ht]
  \centering
  \includegraphics[width=0.5\textwidth]{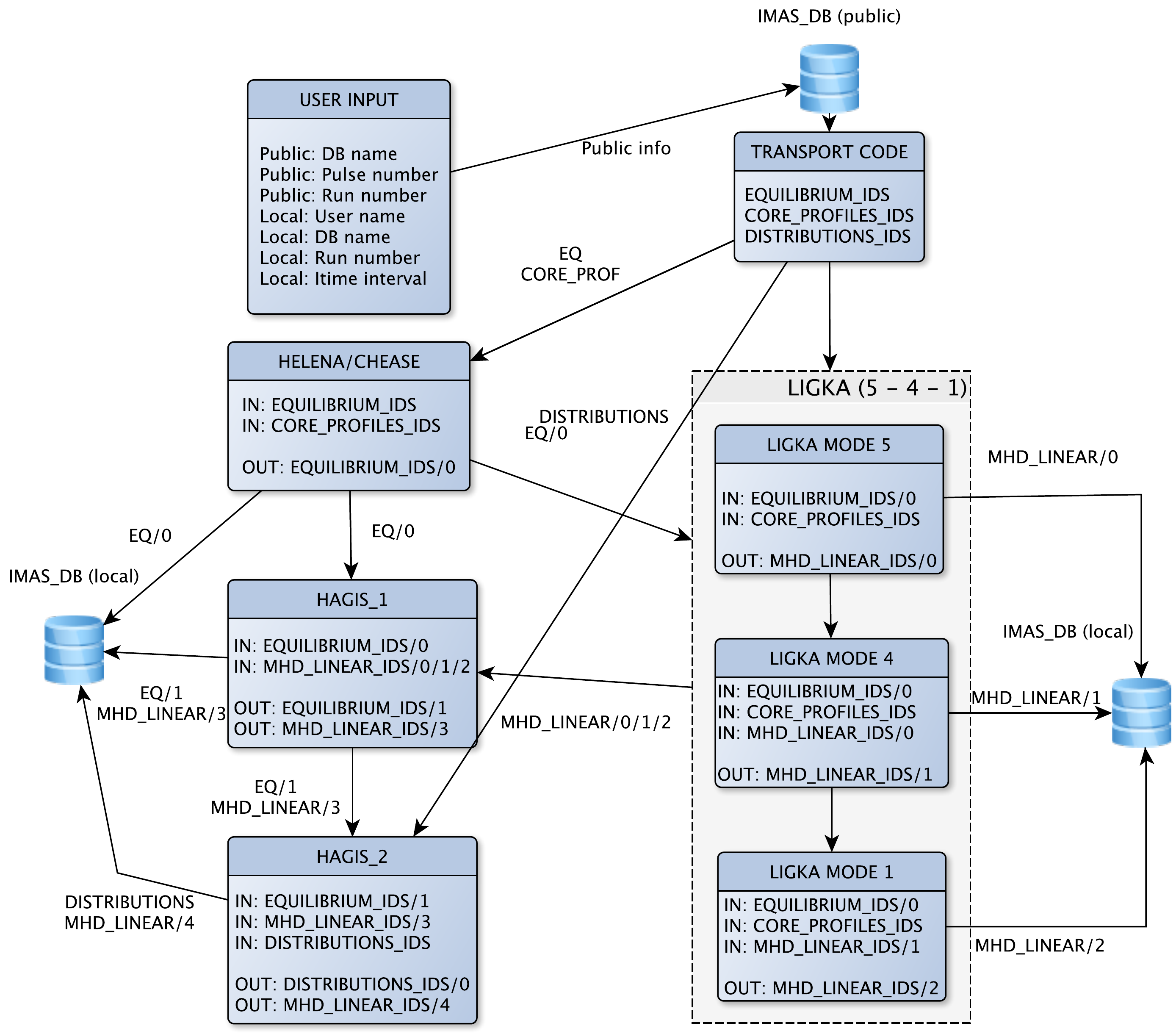}
  \caption{General layout of codes inside the IMAS compliant EP-WF.}
  \label{fig:WF_schema}
\end{figure}
The aim of the WF is to perform an automated linear stability analysis on different time slices of a projected scenario or reconstructed experimental equilibrium. This is achieved by connecting various numerical tools with the data infrastructure (IMAS), facilitating the retrieval/saving of data from/to the database (DB) through a series of IDS files and fast configuration of numerical tools via XML files.
As described in ref. \cite{Imbeaux2015} IMAS relies on IDSs (integrated data structures) that combine a set of well defined physical quantities into the complete description of a complex physical object such as an equilibrium or the linear stability information related to this equilibrium.
In Fig~\ref{fig:WF_schema} one can see how the IDSs are being passed throughout the workflow between the codes and the DB. Input and output of each actor is well defined in terms of IDSs and the static configuration parameters that define the respective simulation are given in XML. In order to facilitate the interaction with the code and the underlying infrastructure, a user interface (GUI) is also available, this is made possible by the standardization of code specific parameters in XML files.

Fig~\ref{fig:WF_schema} explains the how the individual actors are combined:
\begin{enumerate}
  \item Take the equilibrium\_ids from the transport code (given by ASTRA, METIS or DINA-JINTRAC) pass ot to HELENA or CHEASE to perform high-resolution MHD equilibrium calculations. As output they will produce another equilibrium\_ids/0 (*/0 means occurrence 0, one can have several instances of the same IDS in one workflow/database).
  \item Together with equilibrium\_ids/0, core\_profiles\_ids is being taken from the transport code and given as inputs for model 5 of LIGKA (see next subsection). Model 5 performs analytical estimates of TAE (and other AEs such as e.g.\ BAEs, EAEs, RSAEs) and creates as an output mhd\_linear\_ids/0 which in turn it is being saved in the local database.
  \item Next is model 4 of LIGKA which takes as an input equilibrium\_ids/0, core\_profiles\_ids/0 and mhd\_linear\_ids/0 written previously by model 5 and creates a new occurrence of the mhd\_linear\_ids/1. Model 4 returns a local estimate of the linear AE properties, i.e.\ it solves the local kinetic dispersion relation \cite{Lauber2009,Lauber2013,Bierwage2017}.
  \item Finally, model 1 and 2 of LIGKA, the standard global solver can be started. As necessary input IDSs we have: equilibrium\_ids/0, core\_profiles\_ids/0 and mhd\_linear\_ids/1 (output from model 4). Again, it creates a new occurrence of mhd\_linear\_ids/2.
\end{enumerate}

This WF is not only driven by the need to analyze the stability problem with different levels of complexity and speed that is reflected by the different LIGKA models, but also, each LIGKA model profits from the results of the previously run model: the local dispersion relation solver (model 4) needs a good starting guess for the complex contour integration in the frequency plane that is provided by the analytical estimate (model 5). The global solver (model 1/2) in turn uses model 4 results to set up an efficient 'antenna' scan based on the knowledge of the kinetic continuum structure. Consistent mapping of the results of each model into the DB allows us to construct a consistent hierarchy of results and eventually perform uncertainty quantification for each model. Obviously, the modular setting and the usage of standard IMAS data structures facilitates also the usage of other codes than LIGKA under the same umbrella in the future.

In practice, the analytical (model 5) and local solvers (model 4) are used to obtain a fast overview of the scenario before attempting global, more expensive runs.
The global solver (model 1) can then be used to validate - and in most cases - to improve the results obtained by the local runs.

\begin{figure}[htp]
  \begin{minipage}[h]{1\linewidth}
  \centering
  \includegraphics[width=0.7\linewidth]{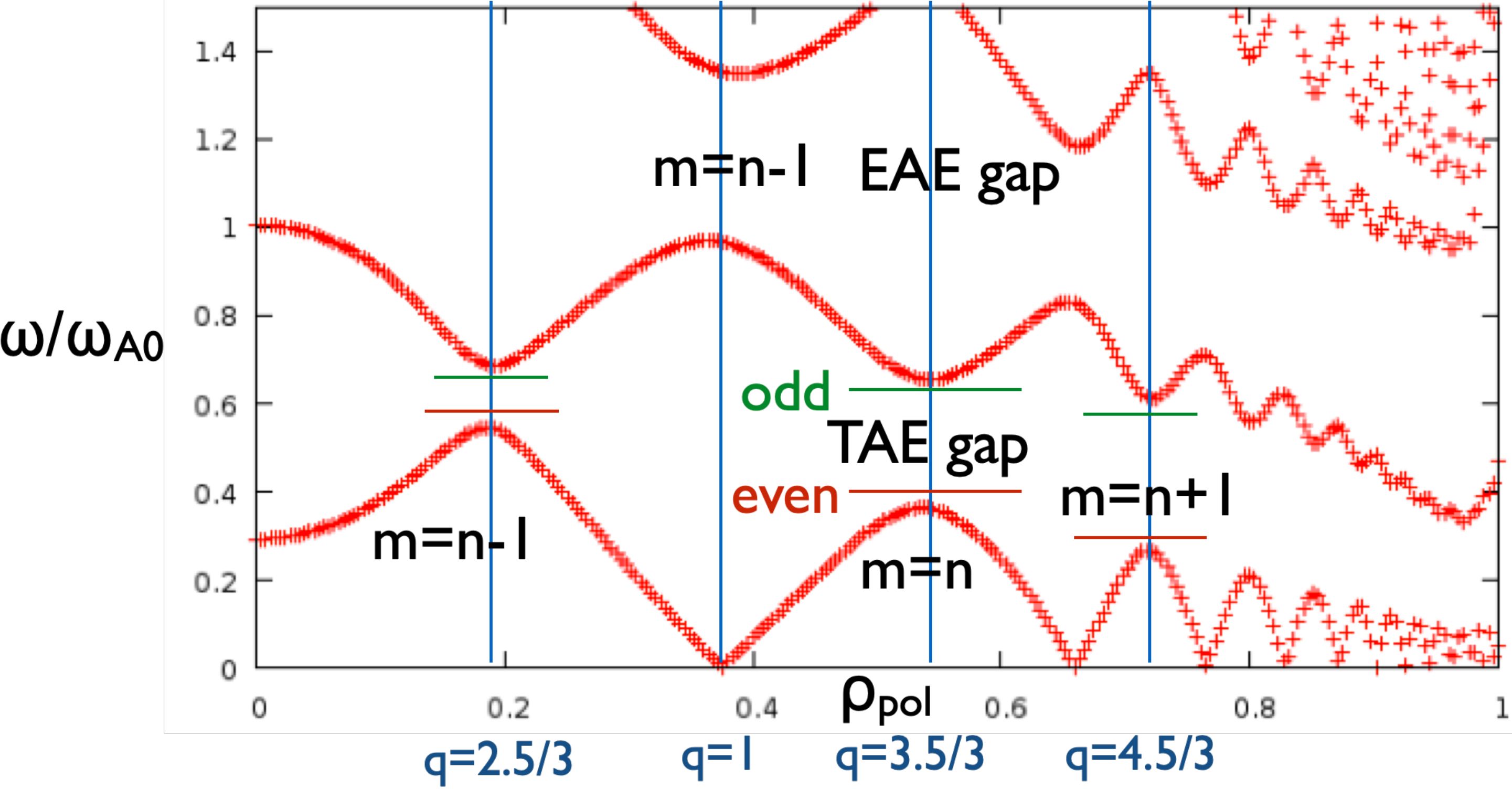}
  \end{minipage}
  \vspace{0.00mm}

  \begin{minipage}[h]{1\linewidth}
  \centering
  \includegraphics[width=0.7\linewidth]{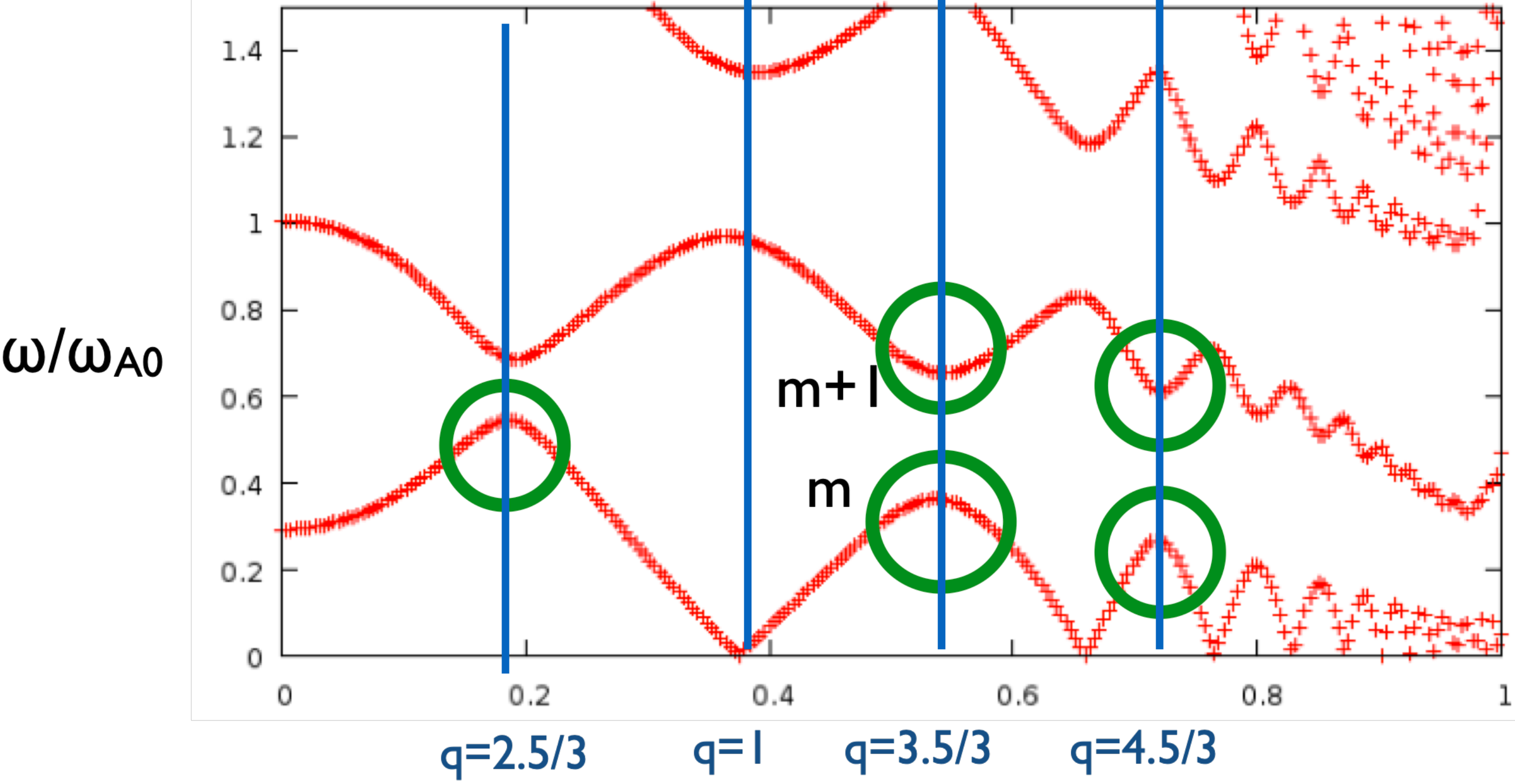}
  \end{minipage}
  \vspace{0.00mm}

\begin{minipage}[h]{1\linewidth}
\centering
\includegraphics[width=0.7\linewidth]{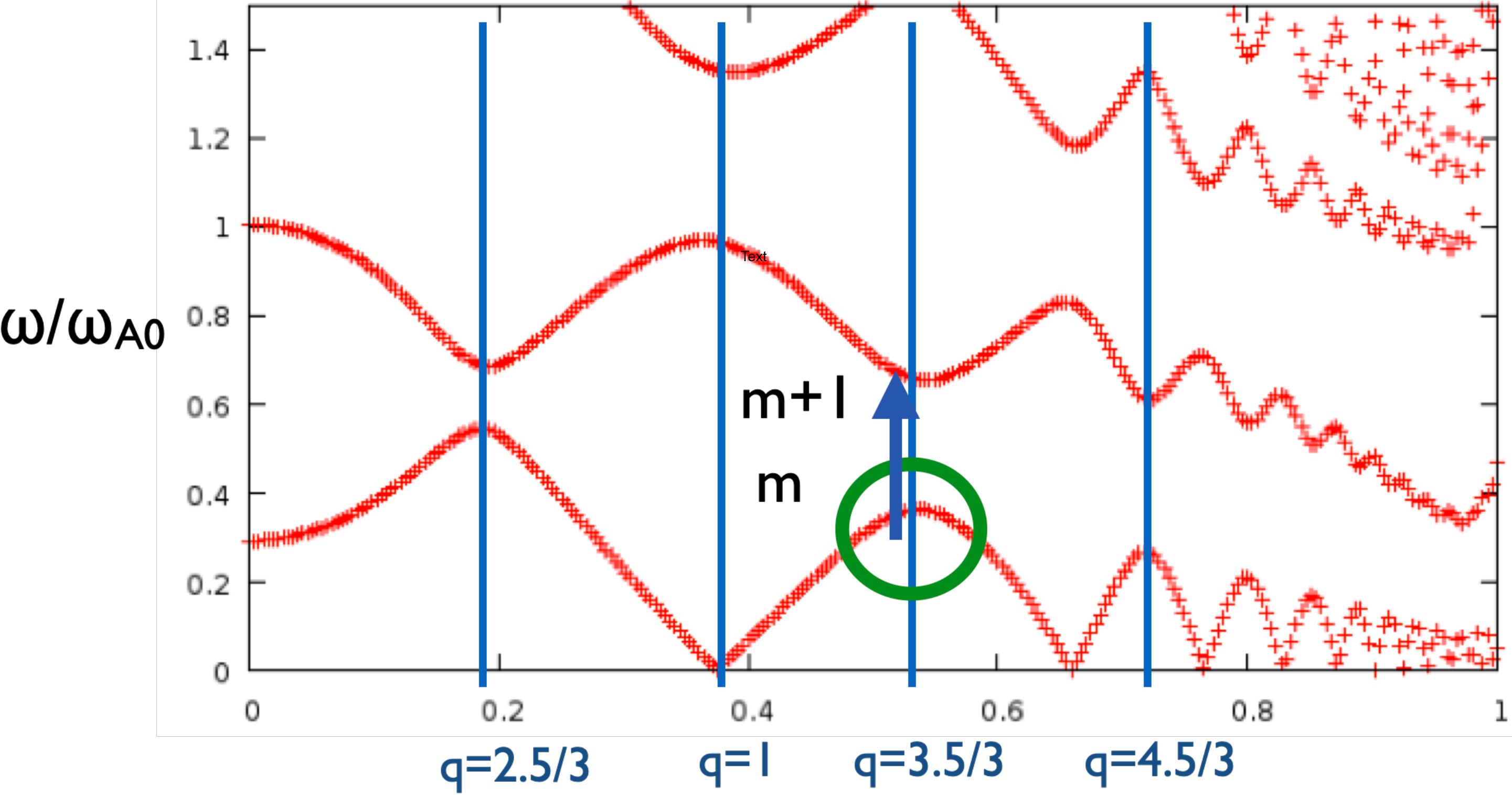}
\end{minipage}

\caption{From top to bottom (model 5/4/1) model explanation.}
\label{fig:models_explained}
\end{figure}

By solving the linearized gyrokinetic equations LIGKA outputs eigenvalues and eigenfunctions such as frequency, damping and mode structures (electrostatic and electromagnetic components, and derived from that also the parallel electric field). Models of LIGKA used in this work are:
\begin{enumerate}
  \item (1 second/mode) Model 5: local analytical estimates of various AEs properties: frequency, estimated mode structure, rational surface, next and previous gap information based on the analytical formulae of ref. \cite{Fesenyuk_2012, Pinches_ITER_2015} (in case we are dealing with AEs related to gaps in the shear \Alfv{} continuum).
  \item (10 seconds/mode) Model 4: based on model 5 results, the local analytical dispersion relation for one AE (n,m-pair) is calculated. The output is the complex frequency of the maximum or minimum of the continuum closest to the AE's rational surface (see Fig.\refeq{fig:models_explained} (middle)), giving a good estimate for ion LD (Landau Damping) and local electron LD. This model determines the starting point for global calculations.
  \item (2-5 minutes/mode) Model 1: based on the results of model 4 (or 3), perform a frequency scan throughout the gap or close to the local extremum (BAE, RSAE) in order to find linear properties of AEs, i.e.\ the location of the global AE in the gap (see Fig.\refeq{fig:models_explained}(last)).
  \item (2-5 minutes/mode) Model 2: uses the same solver as model 1, but model 2 tries to accurately find the phase jump during the frequency scan, such that a more accurate growth/damping rate can be obtained, or the runtime can be reduced by starting the scan close to the AE's frequency as determined by model 1.
  \item (1-10 minutes/mode) Model 6/3: reduced MHD/kinetic spectrum. The details of the kinetic dispersion relation, solution methods and benchmarks ca be found previous papers~\cite{Lauber2009,Bierwage2017}.
\end{enumerate}

In this paper, we focus exclusively on TAEs for simplicity, although also other types of AEs such as RSAEs (reversed shear AEs), BAEs (beta-induced AEs) and EAEs (ellipticity induces AEs) are implemented. Toroidal \Alfv{} Eigenmodes (TAE) are the prototype of all \Alfv{ic} Eigenmodes (AE)~\cite{Cheng1985}. They have their frequency inside the shear \Alfv{} continuum gap and their mode structure is given by coupling of two counterpropagating regular shear \Alfv{} waves. TAEs can be excited through resonance by fast particles and fusion products because their parallel group velocity in toroidal devices is typically of the same order as parallel particle velocity. It is well known that these modes can lead to significant radial particle transport and EP losses \cite{Heidbrink2008,Munoz2006,Schneller2012}.
This workflow is also equipped to work with other numerical tools such as HAGIS ~\cite{Pinches1998} 1 and 2 for non-linear stability of EPs, but this surpasses the scope of this work.

 \section{Scenario description}
\label{sec:scenario}
For the purpose of demonstrating the capabilities of the newly developed workflow, three scenarios were chosen from the many available in the ITER - IMAS database.
The scenarios analyzed in this paper are the baseline Deuterium-Tritium predictive ITER scenarios (generated by different transport codes). The transport data was obtained from running the ASTRA, METIS or DINA-JINTRAC workflows, transport packages that are available at ITER through the Integrated Modelling \& Analysis Suite (IMAS). 

In order to give a clear overview, this chapter is separated into three sub-chapters, each describing one scenario.

\subsection{ASTRA - 131025/34}
This specific ASTRA-generated scenario is time - independent, i.e. only one time slice at 208s in the flat top phase has been exported into IMAS. The plasma composition is a 50:50 Deuterium - Tritium mix. 

\begin{figure}[ht]
  \centering
  \includegraphics[width=0.4\textwidth]{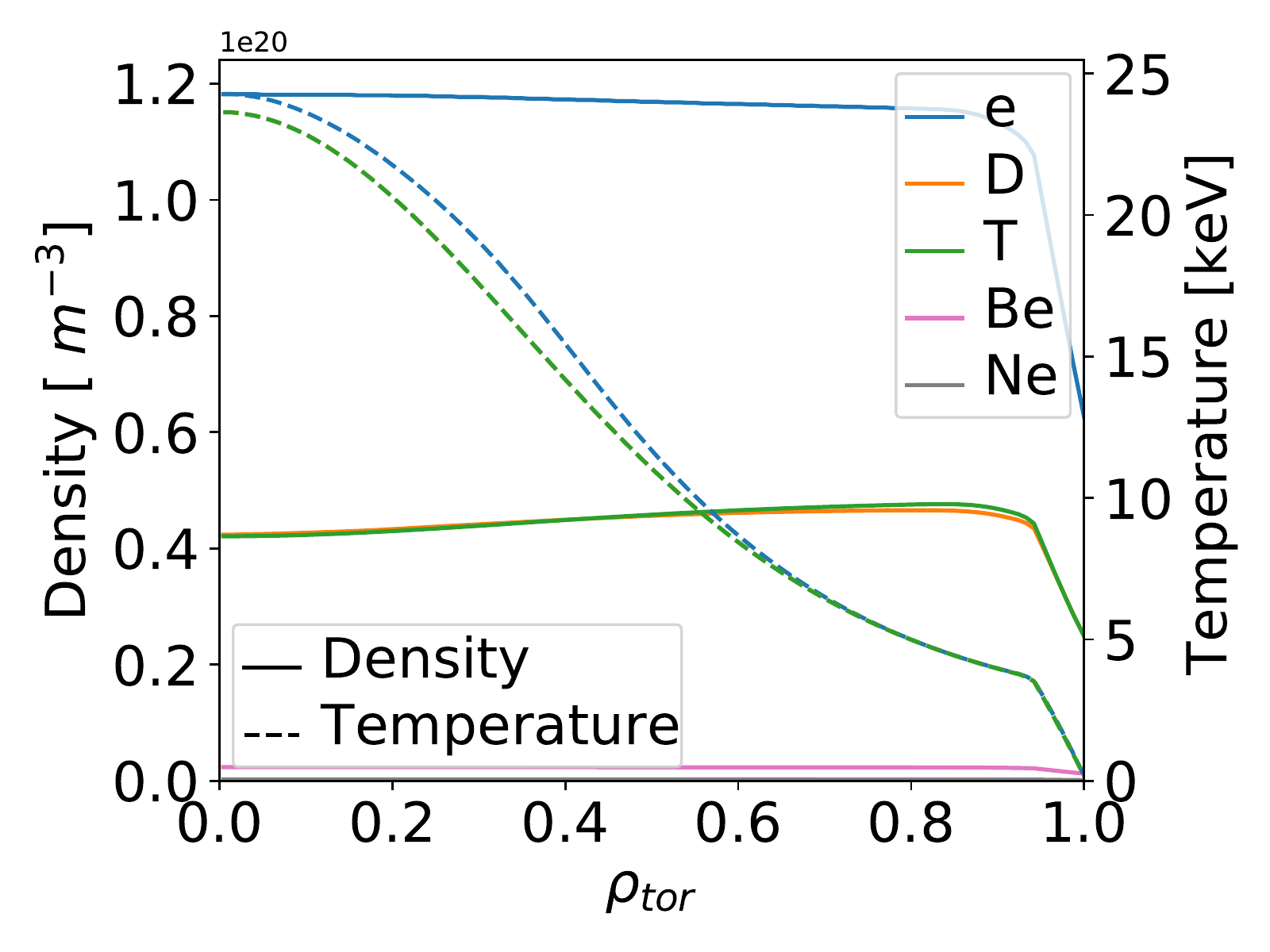}
  \caption{ASTRA scenario density (solid) and temperature (dashed) for electrons (e), deuterium (D) and tritium (T) at $t=208$s.}
  \label{fig:ASTRA_density_plots}
\end{figure}
\begin{figure}[ht]
  \centering
  \includegraphics[width=0.4\textwidth]{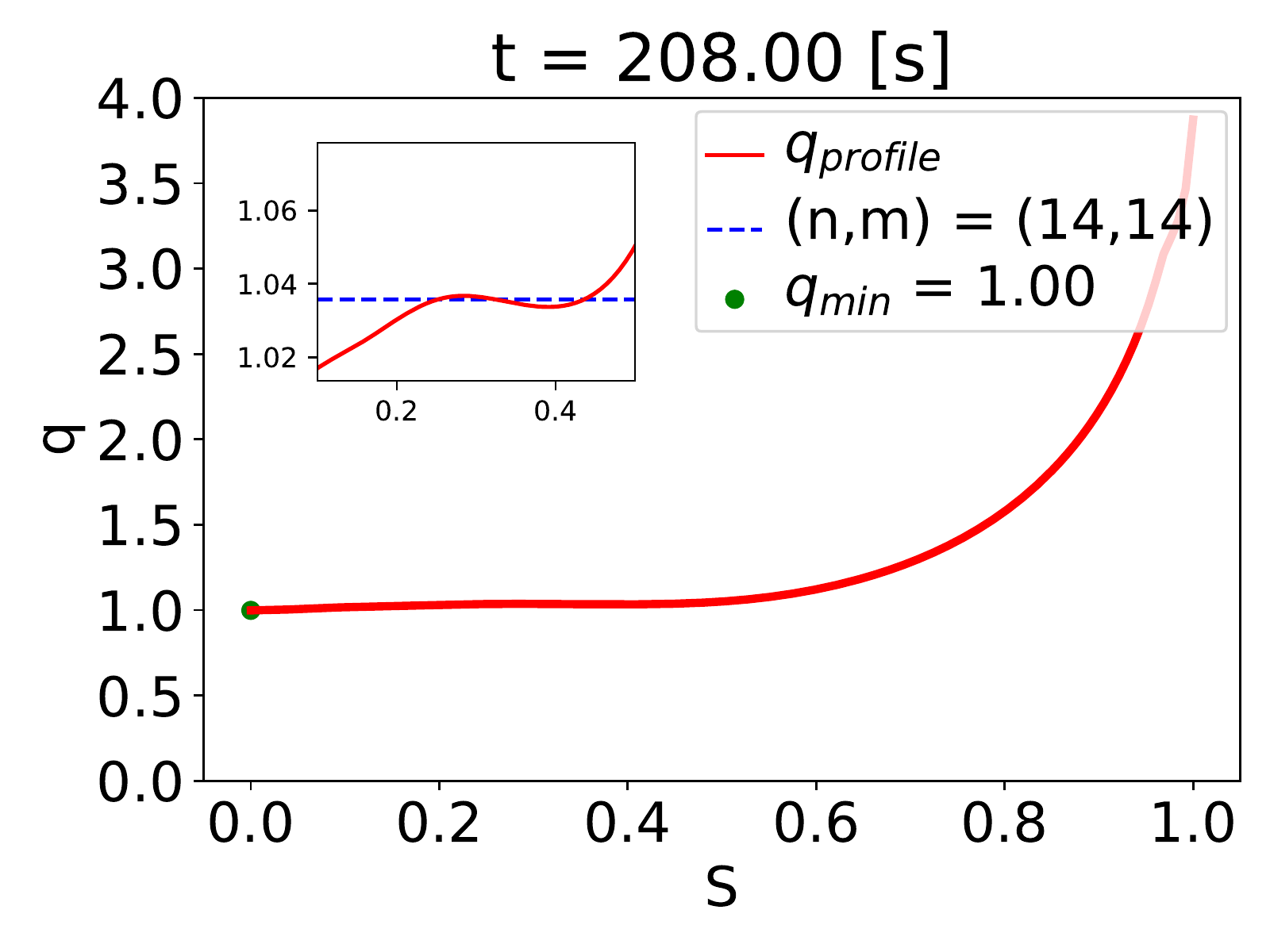}
  \caption{Safety factor profile at $t=208$s. Zoom on the inner part of the tokamak (0.2 - 0.6) to observe that the q profile is not strictly monotonic. With depiction for minimum value of the profile (green). And the rational surface for $n=14$, $m=(14,15)$ TAE (blue). }
  \label{fig:q_profile_astra}
\end{figure}

The electron, deuterium and tritium profiles can be seen in Fig.\refeq{fig:ASTRA_density_plots}. Electrons have a temperature on axis of $T_{e,0} \approx 23$ keV, and $T_{e,p} \approx 4$ keV at the top of the pedestal. For D and T, the on axis temperature $T_{D/T,0}\approx 22$ keV with the temperatures at the pedestal the same as for electrons. We have an almost flat density profile for electrons with $n_{e,0} \approx 1.2 \times 10^{20}$ $m^{-3}$.  For the ions species considered in this scenario (D and T)  the on-axis density is $n_{D/T,0} \approx 0.4 \times 10^{20}$ $m^{-3}$. The profiles peak slightly in the pedestal region $n_{D/T,p} \approx 0.5 \times 10^{20}$ $m^{-3}$ due to the presence of fast ions in core and the assumed impurities profiles (Be, Ne), fulfilling the quasi-neutrality condition.

In Fig. \refeq{fig:q_profile_astra} the safety factor profile (with a zoom showing the values in the inner part of the tokamak and the minimum value of the profile) can be seen. $q_0=1.00$ in this case, meaning that n principle due to $q_{TAE}=(m+0.5)/n$ all TAEs with $n \ge 1$  could be present. Also, the q-profile is not strictly increasing. This can lead to multiple TAE surfaces, as shown by the blue line that indicates $q_{TAE}=(14+0.5)/14=1.036$ for the (n,m) pair (14,14).

\subsection{DINA-JINTRAC - 134173/106}
The second scenario taken into consideration is a time-dependent D-T scenario produced with the DINA-JINTRAC transport workflow. Since it uses a  1.5D core/2D SOL transport model, the profiles can be expected to be more realistic than the 1D METIS model.  Nevertheless, this scenario is still under development and is constantly improved. 

\begin{figure}[ht]
  \centering
  \includegraphics[width=0.5\textwidth]{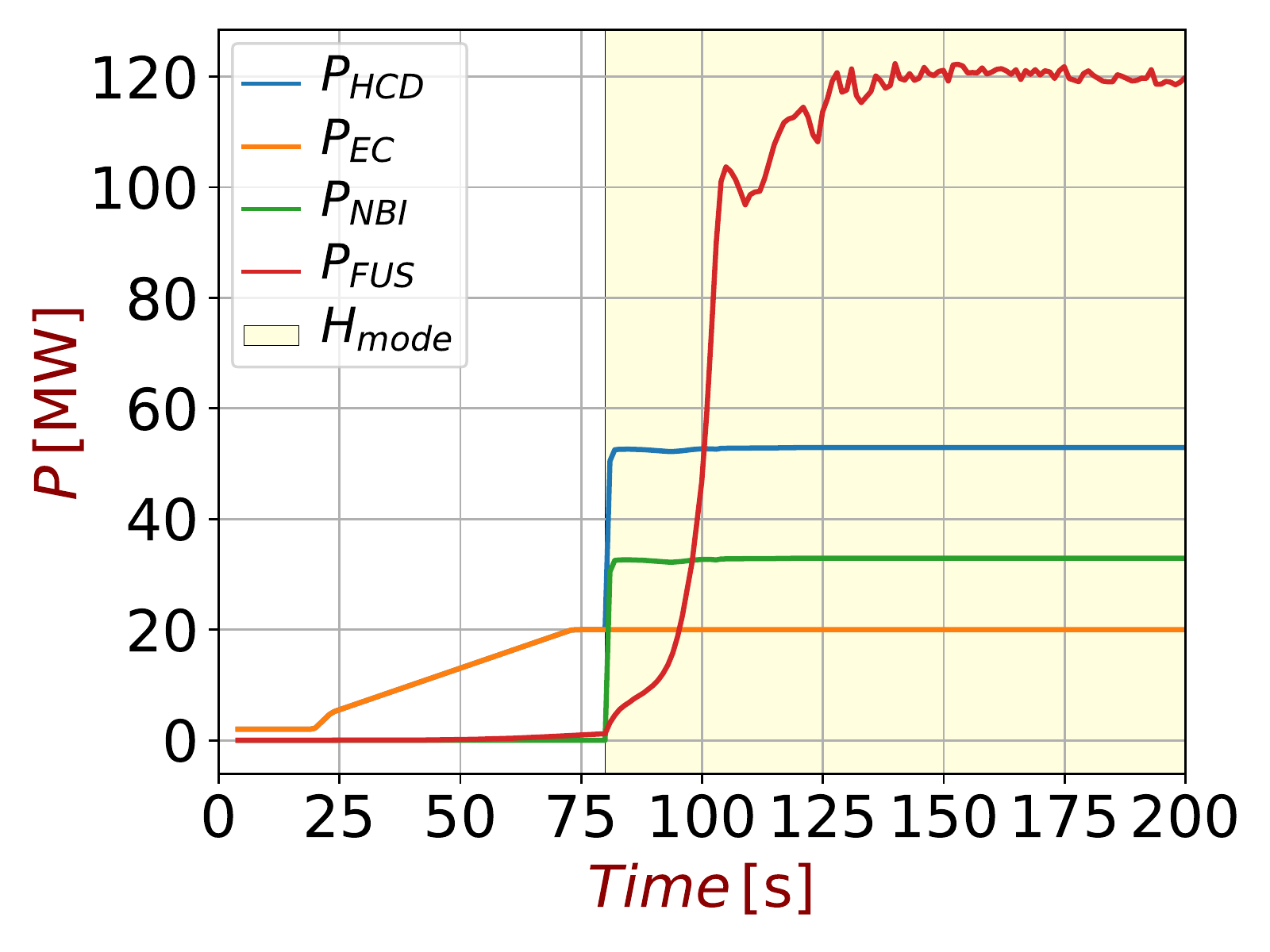}
  \caption{DINA-JINTRAC heating and fusion power evolution until 200s in the scenario.}
  \label{fig:jintrac_fusion_power}
\end{figure}

In Fig.\refeq{fig:jintrac_fusion_power} the fusion power peaks at 150s in the simulation and then slowly decays over the rest of the scenario. Unfortunately, at the time of writhing this paper no fast particle information was available in the transport code output, and thus, no fast particles (EPs) were included in the analysis of this scenario.

\begin{figure}[ht]
  \centering
  \includegraphics[width=0.4\textwidth]{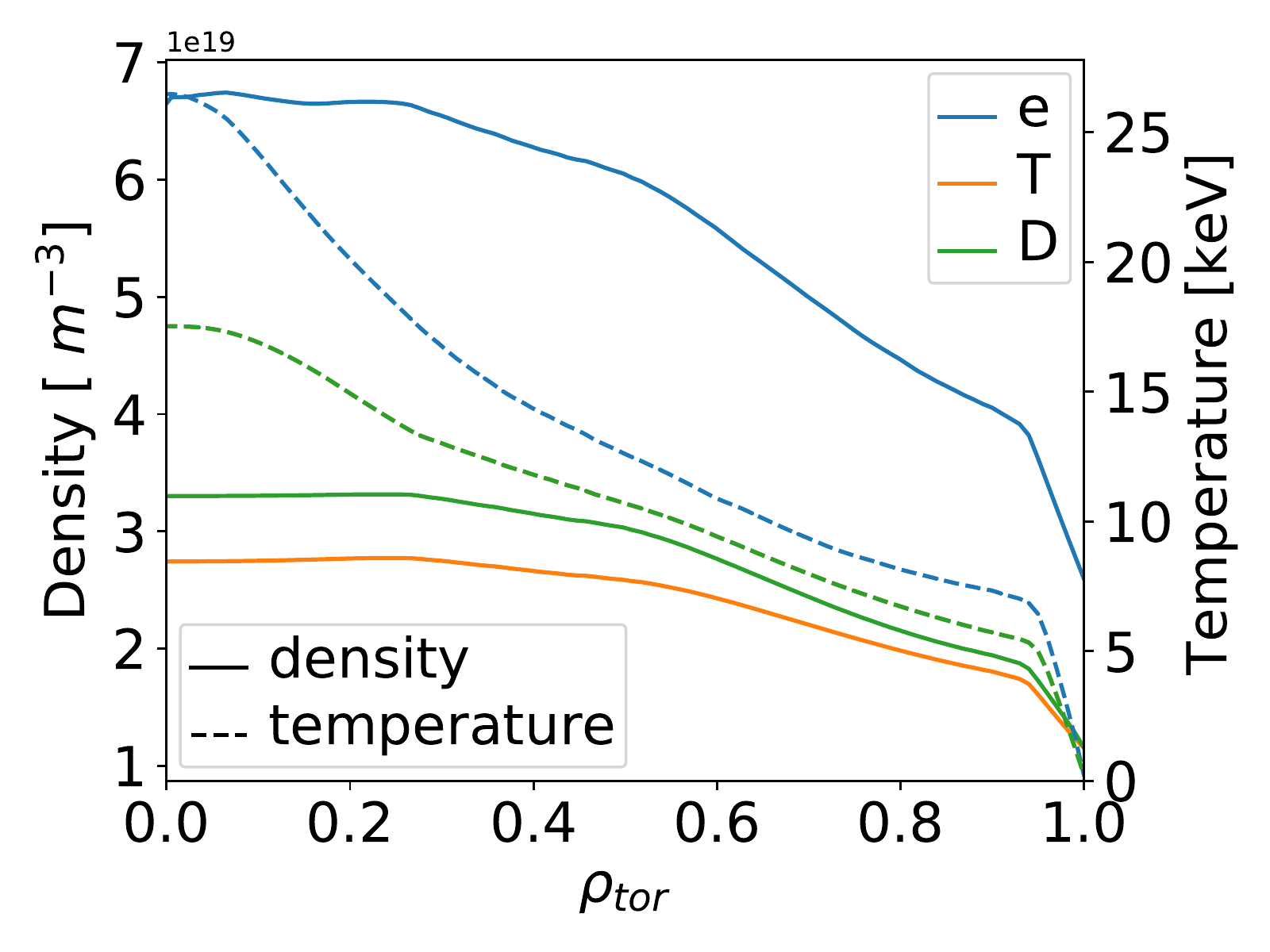}
  \caption{JINTRAC scenario density and temperature for electrons, deuterium and tritium at $t=97.94$s.}
  \label{fig:JINTRAC_density_plots}
\end{figure}

In Fig.\refeq{fig:JINTRAC_density_plots} density and temperature for electrons and background ions are shown. The temperature profiles, with dashed lines are as follows: $T_{e,0} \approx 26$ keV, $T_{e,p} \approx 6$ keV, $T_{D/T,0}\approx 16$ keV and $T_{D/T,p}\approx 5$ keV. The densities are: $n_{e,0} \approx 6.8 \times 10^{19}$ $m^{-3}$ for electrons, $n_{D,0} \approx 3.2 \times 10^{19}$ $m^{-3}$ for deuterium and $n_{T,0} \approx 2.8 \times 10^{19}$ $m^{-3}$ for tritium.

\subsection{METIS - 130012/02}
The third scenario, is a time-dependent scenario generated by METIS. It is again a D-T plasma, the baseline ITER scenario for $Q=10$, and thus, one of the most important scenarios for fusion at ITER.

\begin{figure}[ht]
  \centering
  \includegraphics[width=0.4\textwidth]{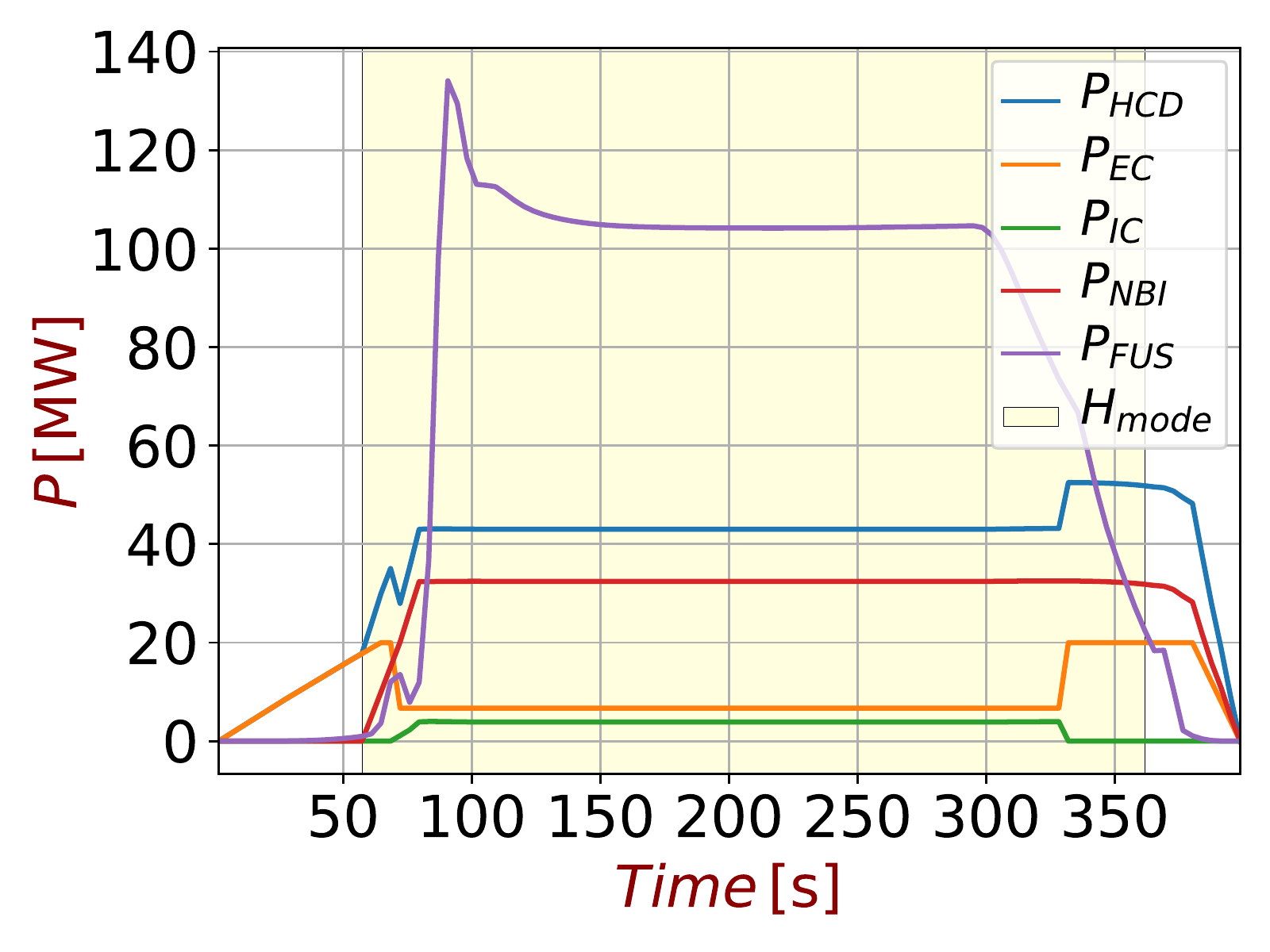}
  \caption{METIS scenario heating and fusion power evolution. Fusion Power peaks at 90s.}
  \label{fig:fusion_metis}
\end{figure}

\begin{figure}[ht]
  \centering
  \includegraphics[width=0.4\textwidth]{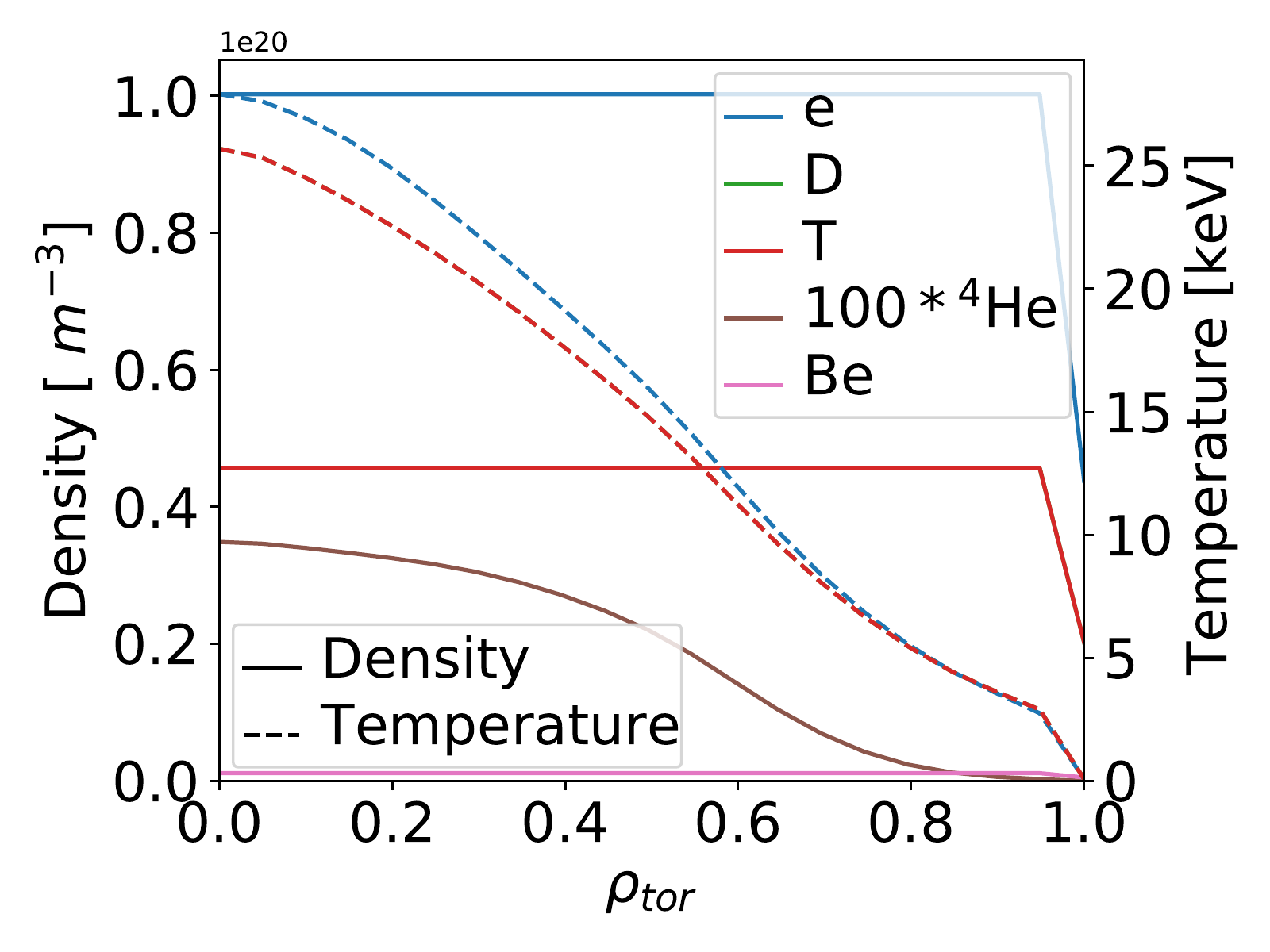}
  \caption{METIS scenario density and temperature for electrons, deuterium, tritium and alpha particles at $t=90$s.}
  \label{fig:density_profile_metis}
\end{figure}

\begin{figure}[ht]
  \centering
  \includegraphics[width=0.4\textwidth]{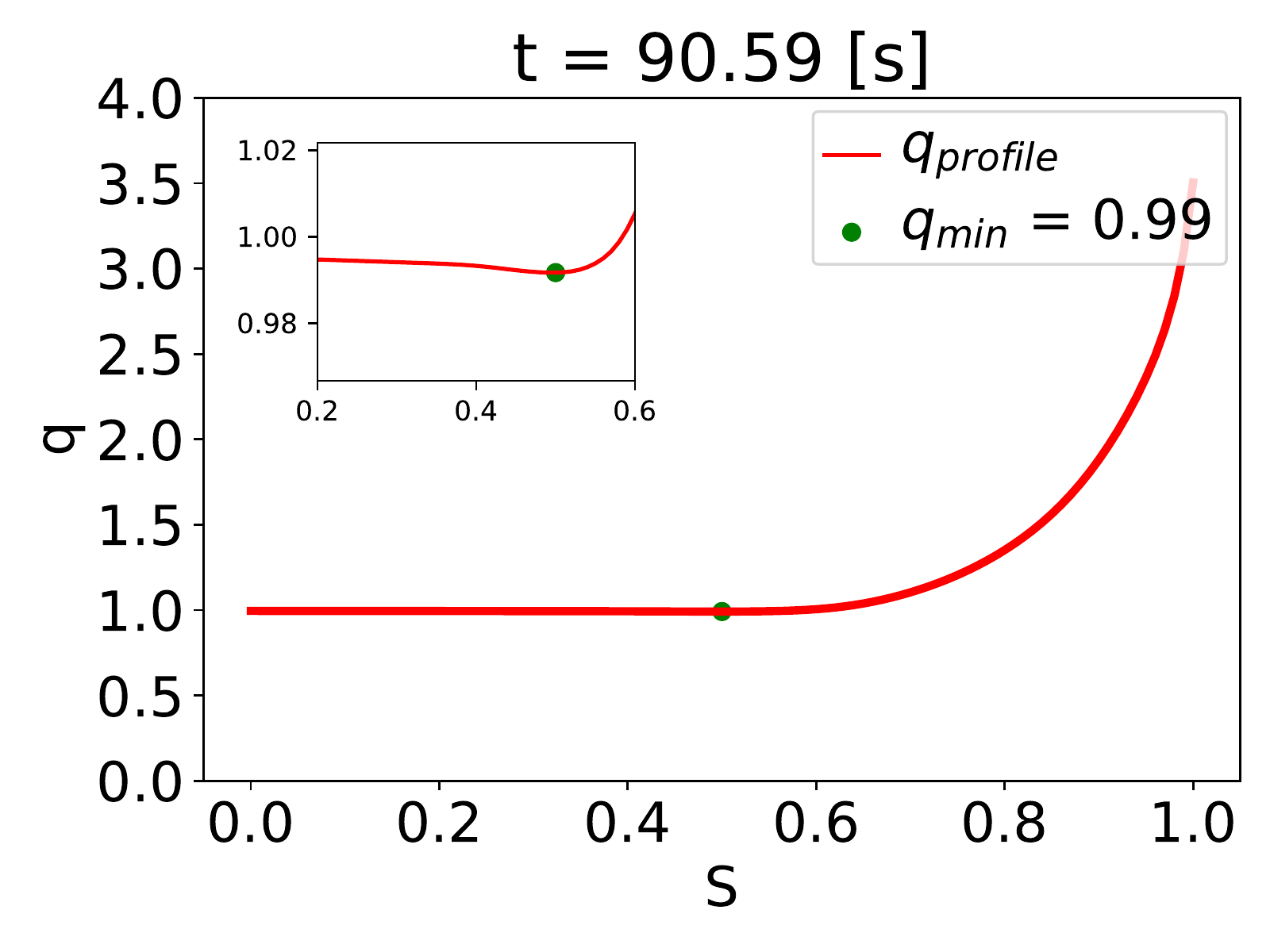}
  \caption{Safety factor profile at $t=90.59$s. Zoom on the inner part of the tokamak. With depiction for minimum value of the profile.}
  \label{fig:q_profile_metis}
\end{figure}

In Fig.\refeq{fig:fusion_metis} the evolution of  heating and fusion powers  for the METIS scenario are shown. Of special interest for our study is the $P_{FUS}$ (purple line). It peaks at approximately $90$ s with $P_{FUS} \approx 130$ MW.  This evolution allows us to test and validate the energetic particle part of the EP-WF.

Fig.\refeq{fig:density_profile_metis} shows, like before, the density and temperature profiles this time at $t\approx90$s i.e. at the peak of $P_{FUS}$. For electrons, the temperatures on axis and at the pedestal peak at $T_{e,0} \approx 27$ keV, and $T_{e,p} \approx 4$ keV. For background ions (D and T in this case) the temperature profile is the same with $T_{D/T,0}\approx 25$ keV at the axis, and same as the electrons at the pedestal. The electron and background ions densities are mostly flat, peaking at $n_{e,0} \approx 1 \times 10^{20}$ $m^{-3}$ for electrons and $n_{D/T,0} \approx 0.45 \times 10^{20}$ $m^{-3}$ for background ions. In this scenario, we have alpha particles from the fusion process. Their density is also shown in the same figure ($100\times$), peaking on axis at $n_{He4,0} \approx 0.38 \times 10^{18}$ $m^{-3}$. 

In Fig.\refeq{fig:q_profile_metis} we have the safety profile at $t=201.96$ s for this scenario and the minimum value of this profile is $q_{min} = 1.06$ depicted by a green dot in the figure.

  \section{Results}
\label{sec:results}
\subsection{ASTRA - 131025/34}
In this time independent scenario, we perform a scan over a wide array of TAEs ($n=1\dots 35$ and $m=n\dots n+4$) for the given time slice in order to check if all WF elements (including the kinetic multi-species part) are behaving as expected. For this case we include only thermal background particles, \ie{} electrons and a 50:50 D-T mix satisfying quasineutrality at each radial slice.

\begin{figure}[ht]
  \centering
  \includegraphics[width=0.4\textwidth]{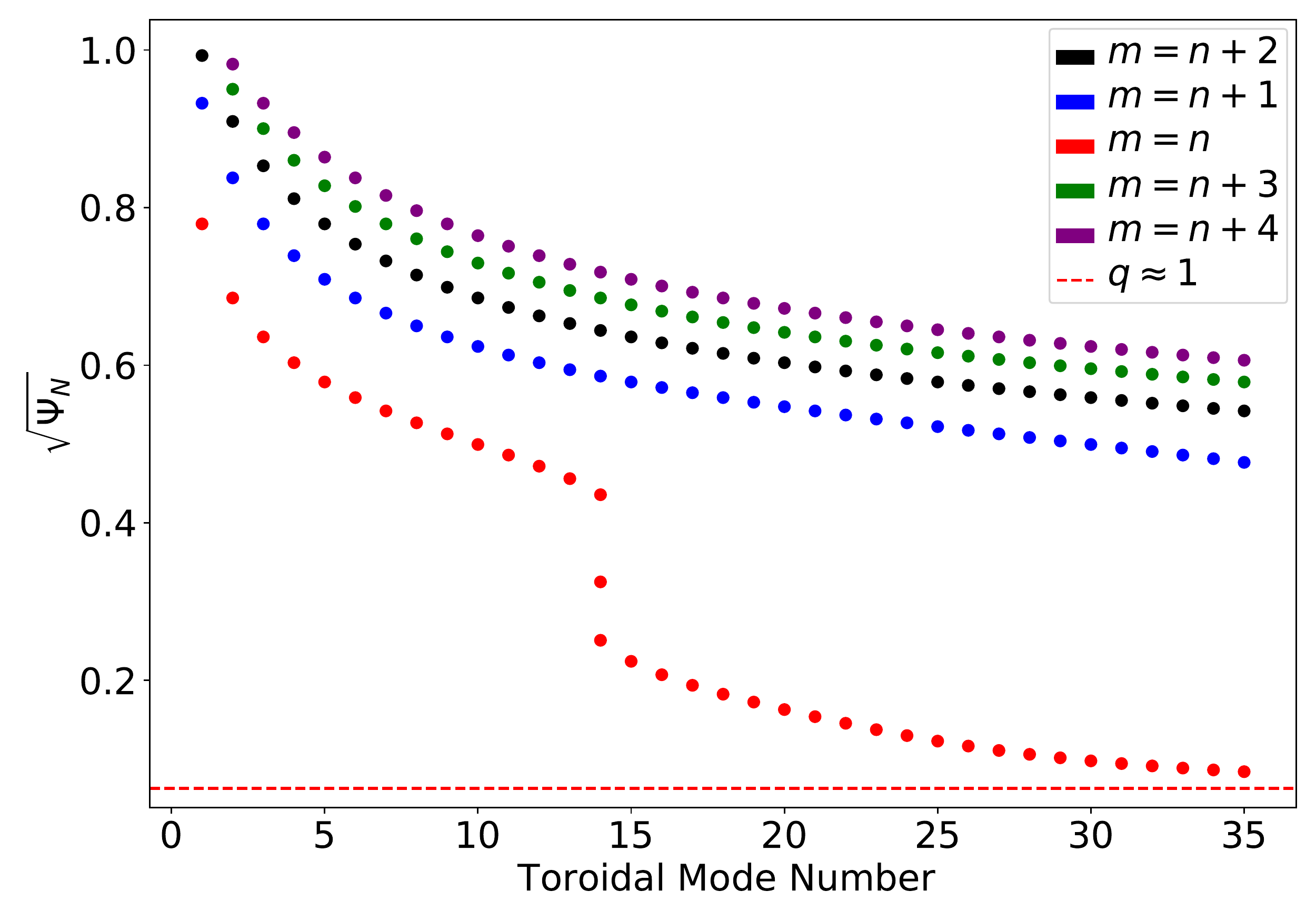}
  \caption{Radial location of modes with toroidal mode number ($n$) between 1 and 35 for $m=n$ to $m=n+4$. For reference, the $q\approx1$ position is depicted with the red dotted line.}
  \label{fig:radial_location_astra_all}
\end{figure}

In Figure \ref{fig:radial_location_astra_all} the radial location of the modes over the whole radial domain is presented. Also the $q\approx1$ rational surface is marked as red line. LIGKA model 5 uses as analytical estimate the TAE condition $q=\frac{m+\frac{1}{2}}{n}$. Also more sophisticated formulae based on \cite{Fesenyuk_2012} are implemented, taking into account elongation and $\beta$ corrections (not shown here). However,  it turns out that the simplest expression gives a good estimate for the center of the TAE gap (that is needed for the subsequent analysis) in all geometries and mode numbers ranges investigated so far. Part of the  modes can be typically discarded in the following analysis (\ie{} model 4 and 1) based on their  position relative to the EP gradient:
\begin{figure}[ht]
  \centering
  \includegraphics[width=0.4\textwidth]{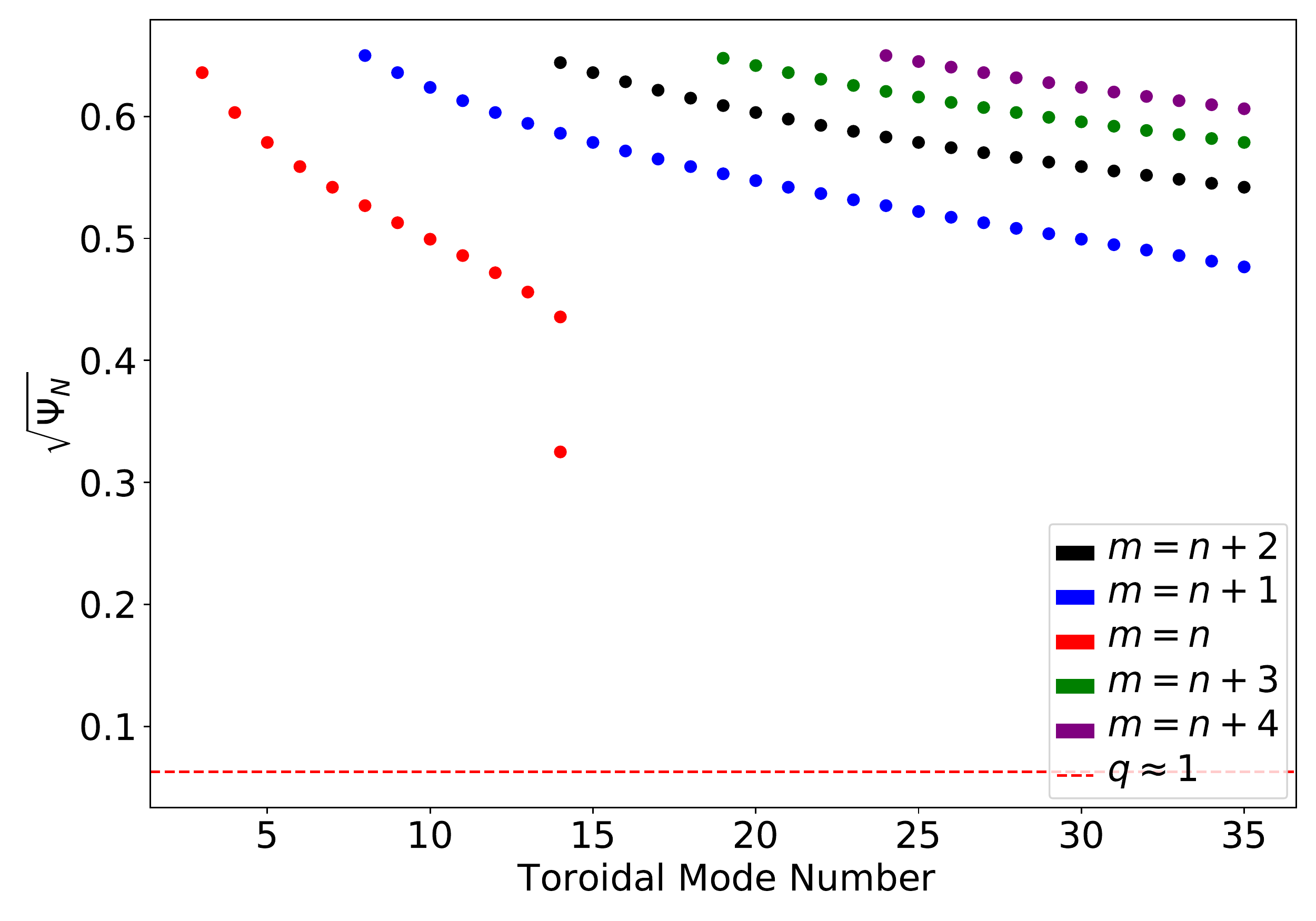}
  \caption{Radial location of modes with toroidal mode number ($n$) between 1 and 35 for $m=n$ to $m=n+4$. This has now been restricted between $0.3\dots 0.65$.}
  \label{fig:radial_location_astra_restricted}
\end{figure}
the alpha particle gradient is significant only  in the region $0.3\dots0.65$, thus  the modes that do not fall in this range were excluded, since they are expected to be linearly stable. The radial range of interest can also be conveniently defined via the input files (xml data structure, editable directly, or via a gui), allowing the user to have a  flexible control over the considered mode spectrum.  Figure \ref{fig:radial_location_astra_restricted} shows the modes that remain.


\begin{figure}[ht]
  \centering
  \includegraphics[width=0.5\textwidth]{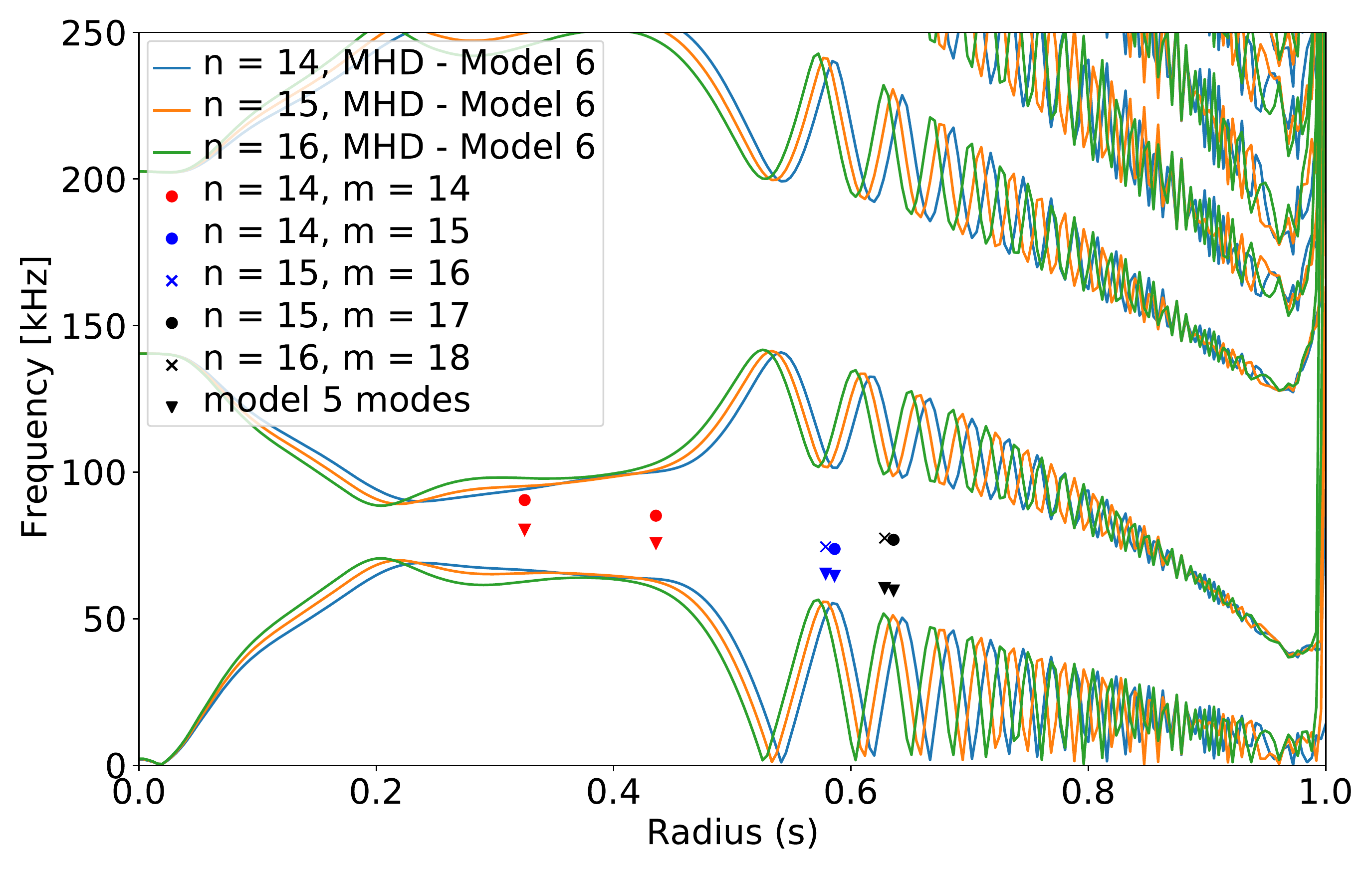}
  \caption{Local continuum for $n=[14,15,16]$ with selected modes position depicted on the same graph. The modes have the same color scheme as the ones in the above analysis on the radial location. Triangles are the simple analytical estimates, the dots global results as obtained with model 1.}
  \label{fig:continuum_modes_astra}
\end{figure}

\begin{figure}
  \centering
  \begin{subfigure}[b]{0.23\textwidth}
      \includegraphics[width=\textwidth]{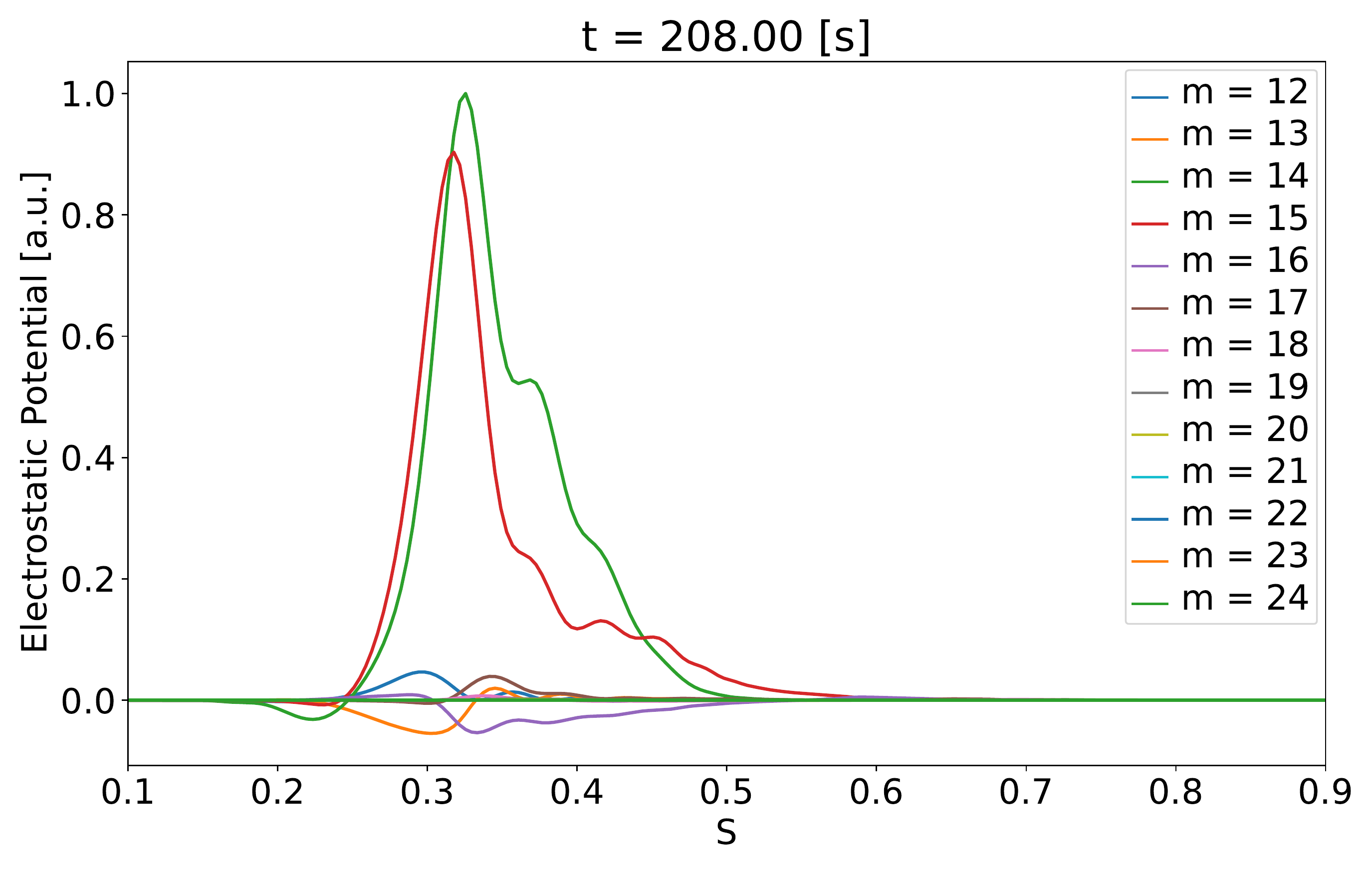}
  \end{subfigure}
  \hspace{0.00mm}
  \begin{subfigure}[b]{0.23\textwidth}
      \includegraphics[width=\textwidth]{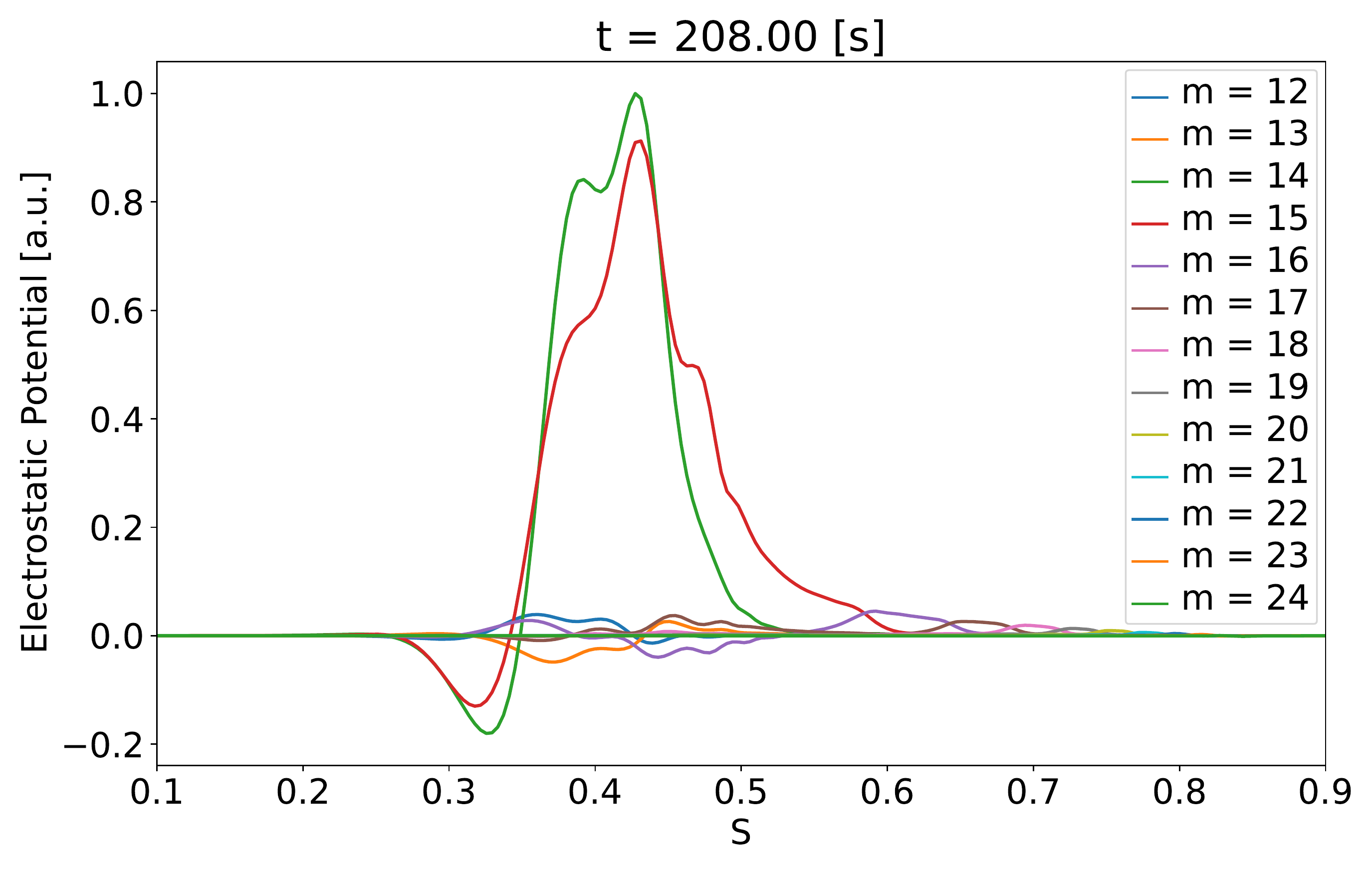}
  \end{subfigure}
  \\
  ~ 
  \begin{subfigure}[b]{0.5\textwidth}
      \includegraphics[width=\textwidth]{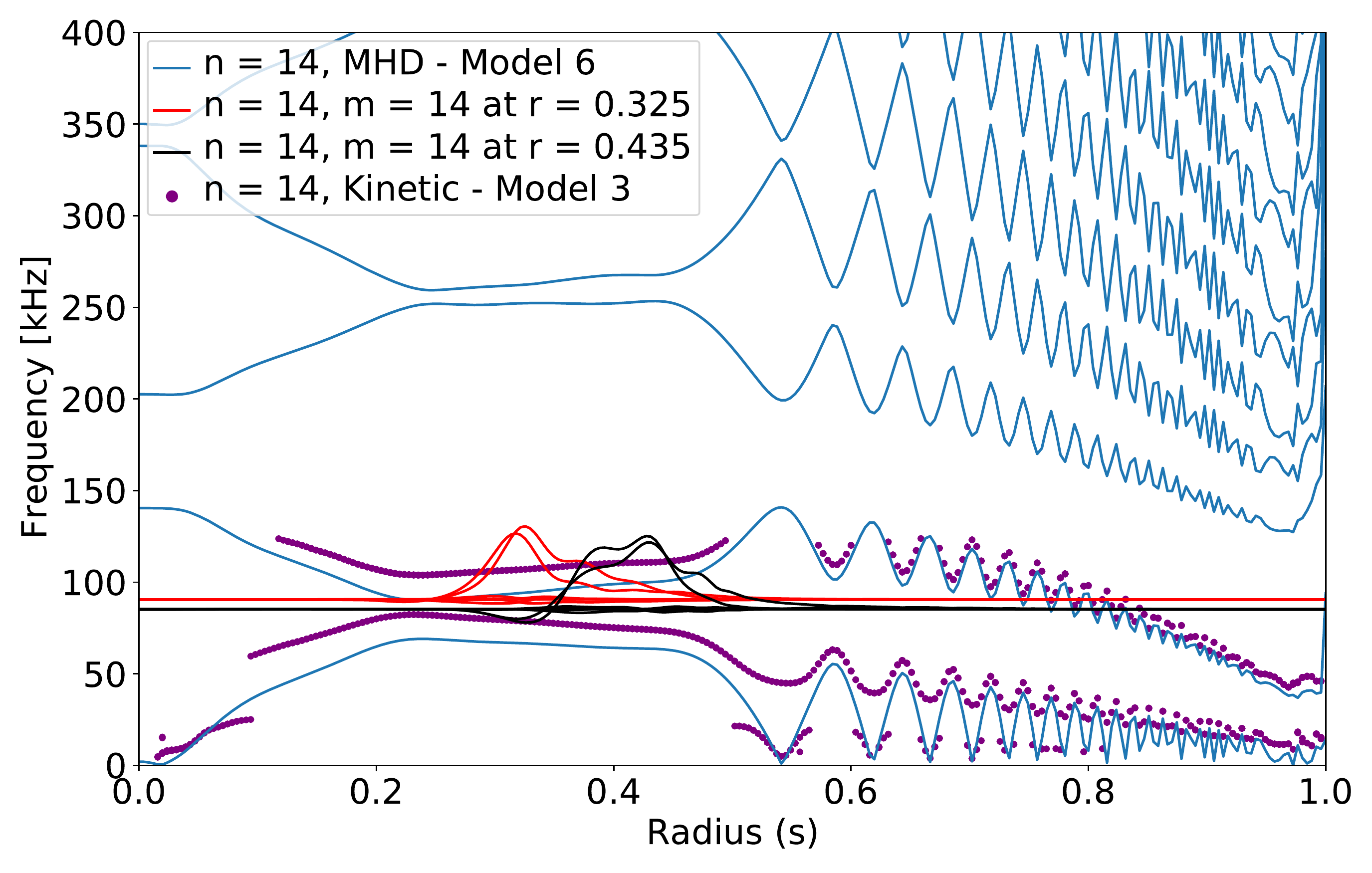}
  \end{subfigure}
  \caption{Top 2 figures contain the real part of the electrostatic potential of the $n=14$ $m=(14,15)$ modes found after restricting our domain. Underneath, is their location on the local continuum (model 6) and kinetic continuum (model 3).}
  \label{fig:ASTRA_EFs_continuum}
\end{figure}
For all selectred toroidal mode numbers the shear Alf\'ven wave continumm (SAW) can be calculated (model 6), as shown in figure \refeq{fig:continuum_modes_astra}) for $n=14,15,16$. All branches of TAEs in the radial range discussed above are added. It can be seen that the simple analytical guesses (model 5) lie typically in the TAE gap, as expected. The two solutions for n=14 are shown (see figure \refeq{fig:q_profile_astra}), separated in frequency by the changes in the local \Alfv{} velocity. Now we calculate the global solutions with model 1.
In Fig.\refeq{fig:ASTRA_EFs_continuum} (round dots) the normalized mode structures (electrostatic potential) for the two modes $n=14$, $m=(14,15)$ found at different radial location are displayed (as explained before). By taking these structures and adding  them to the the continuum, (see Fig.\refeq{fig:ASTRA_EFs_continuum}) one can see that they intersect the reduced SAW continuum. This is not surprising since the global solutions include various kinetic effects not accounted for in the reduced MHD model, in particular the diamagnetic upshift of the continua in regions of high plasma pressure \cite{Lauber2009}. In Fig.\refeq{fig:ASTRA_EFs_continuum} last subfigure, the local kinetic continuum -model 3- is added. For a detailed description of this local solution of the kinetic dispersion function see \cite{Bierwage2017}. From the figure one can see that using the kinetic model consistently in both local and global calculations, the TAEs are, as expected, situated in the kinetic TAE gap, avoiding local continuum damping. It becomes appearent from this discussion, that the kinetic corrections for these ITER scenarios and mode number ranges are significant and cannot be neglected as often done in present-day experiments where both the plasma beta and the excited TAE mode numbers are typically much lower.
\newpage

\subsection{DINA-JINTRAC - 134173/106}
In this subsection, the previously shown capabilities of the workflow are used to analyze the ramp-up phase of a DINA-JINTRAC scenario simulation (134173/106).
\begin{figure}[ht]
  \centering
  \includegraphics[width=0.4\textwidth]{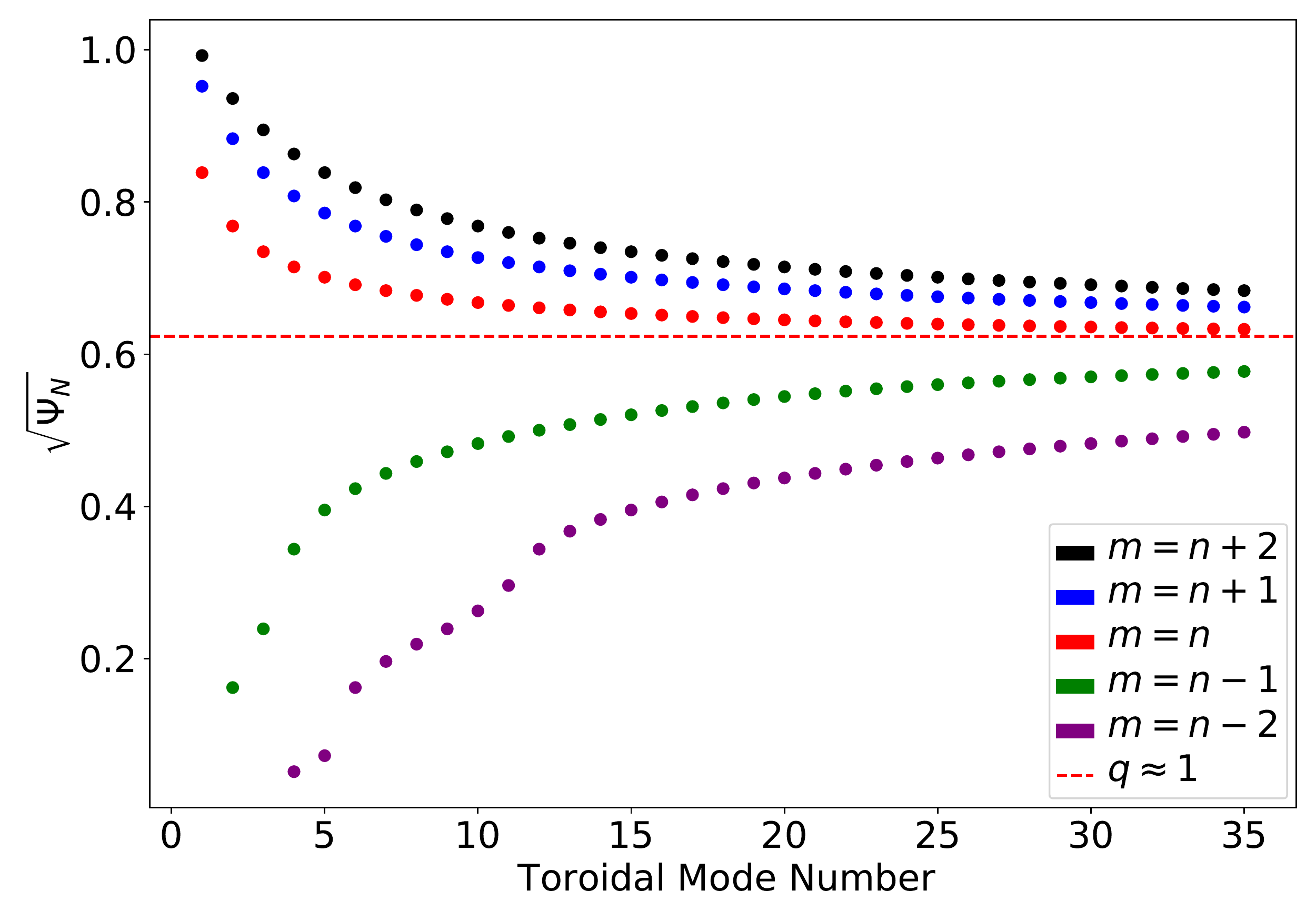}
  \caption{Radial location of modes with toroidal mode number ($n$) between 1 and 35. Poloidal mode numbers are from $m=n-2$ to $m=n+2$. $q\approx 1$ location is depicted with a red line for reference. This run is performed only at $t=97.93$s .}
  \label{fig:radial_position_jintrac}
\end{figure}

Similarly to what has been shown in the last section, a run was performed at one time point ($t=97.93$s), using model 5. For this run,  TAEs with $n=1$ to $n=35$ with $m=n-2$ to $m=n+2$ as main poloidal harmonics were considered. Fig.\refeq{fig:radial_position_jintrac} shows the radial position of these modes together with the $q\approx 1$ rational surface.  As expected, modes with low toroidal mode numbers are shifted away from $q= 1$ towards the core, when the poloidal mode numbers are lower ($m = n-2$) or towards the edge when $m=n+2$ according to $q_{TAE}=\frac{m+\frac{1}{2}}{n}$.

\begin{figure}[ht]
  \centering
  \includegraphics[width=0.4\textwidth]{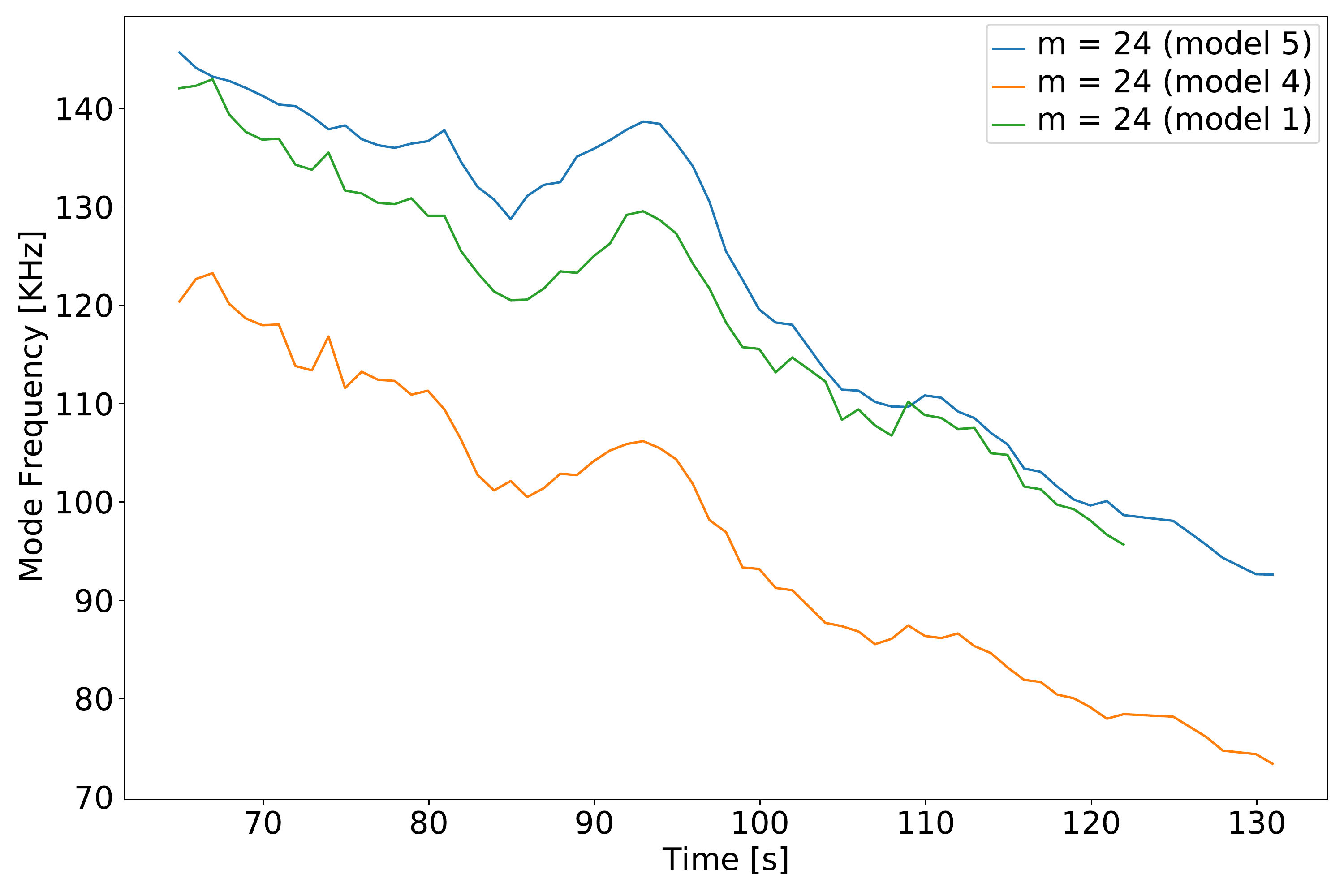}
  \caption{Frequency for the $n=25$, $m=24$ TAE during the ramp-up phase of 134173/106}
  \label{fig:JINTRAC_freq}
\end{figure}

Next, a  specific TAE with $(n,m)=(25,24)$ was tracked during the ramp-up phase of 134173/106. This mode was chosen based on the fact that $q_{min}$ is slightly below 1 in most of the time points, meaning that the TAE with $(n, m=n-1)$ is present throughout the whole ramp-up and flat top phase.

Fig.\refeq{fig:JINTRAC_freq} shows the frequency evolution. Plotted are the different models used by LIGKA. Starting with model 5 (blue line), continuing with the local solver, model 4 (orange line) and finally the global solver, model 1 (green line). Performing  convergence tests as described above we found that the range of poloidal harmonics [$-4$,$+32$] is necessary to study this particular part of the scenario. Features from the computations regarding frequency of the mode can be seen here as well in  addition to the general \Alfv{} scaling trend: due to the rising density, the frequency decreases due to $f_{TAE}\approx \frac{1}{\sqrt{n}}$. With model 5 resulting in the highest frequency (middle of the gap), model 4 the closest continuum (downwards, lower frequency) and finally model 1 determines the exact location of the TAE within the gap via a global analysis.

\begin{figure}[ht]
  \centering
  \includegraphics[width=0.4\textwidth]{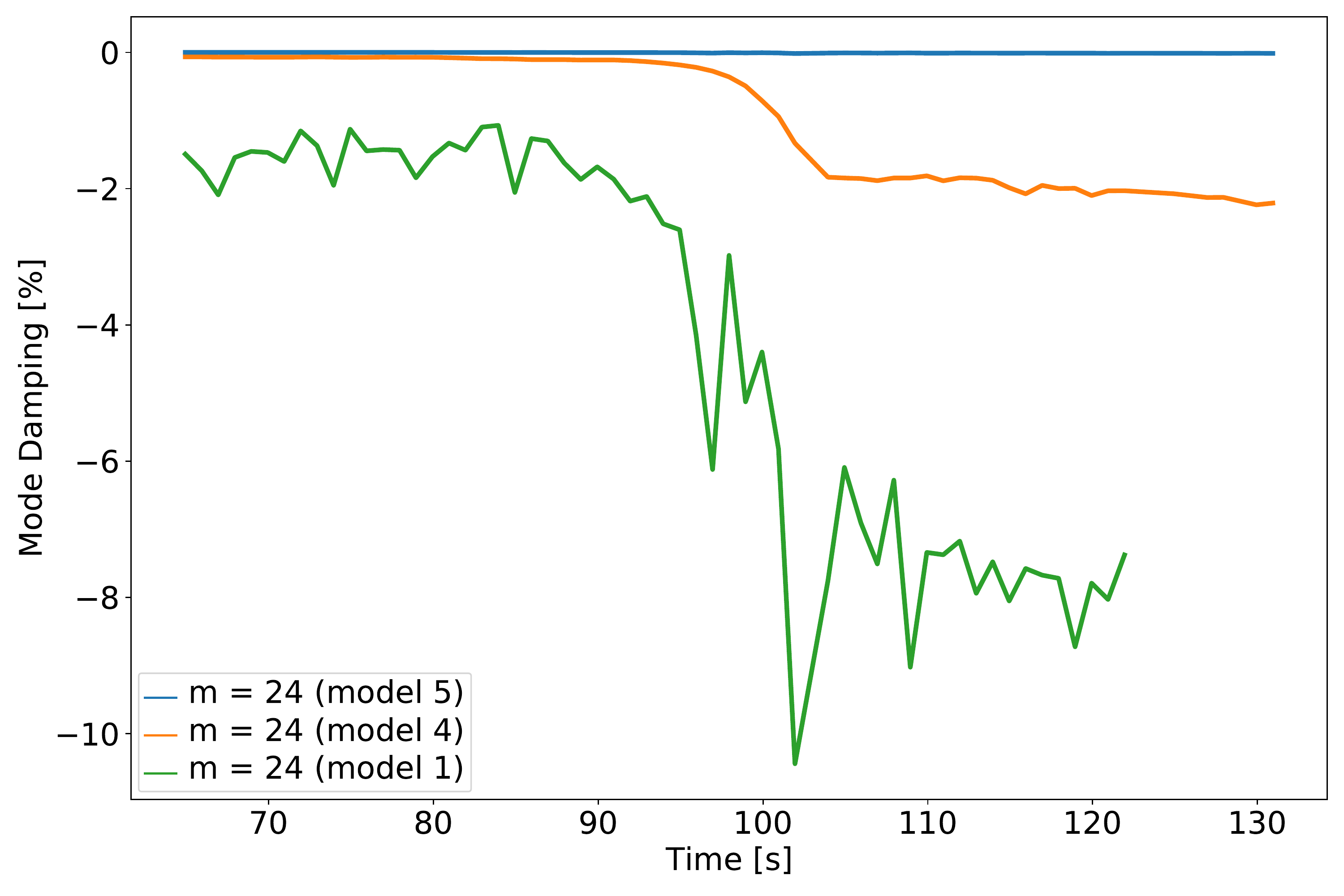}
  \caption{Damping rate for $n=25$, $m=24$ mode along the ramp-up of the scenario.}
  \label{fig:JINTRAC_growth}
\end{figure}

Fig.\refeq{fig:JINTRAC_growth} shows the damping rate of the TAE, during the same period of time (ramp-up). Again, a considerable difference between models can be noticed, with model 1 being the most reliable, showing clearly the the local damping rate under-estimates the global one considerably, as explained above. It can be seen that the chosen TAE is strongly damped in the flat top, whereas during the ramp-up phase the damping rate is much smaller. Clearly this motivates future detailed studies including the drive from the neutral beams and the the alpha particle drive that were not included in this calculation.

\begin{figure}[ht]
  \centering
  \includegraphics[width=0.5\textwidth]{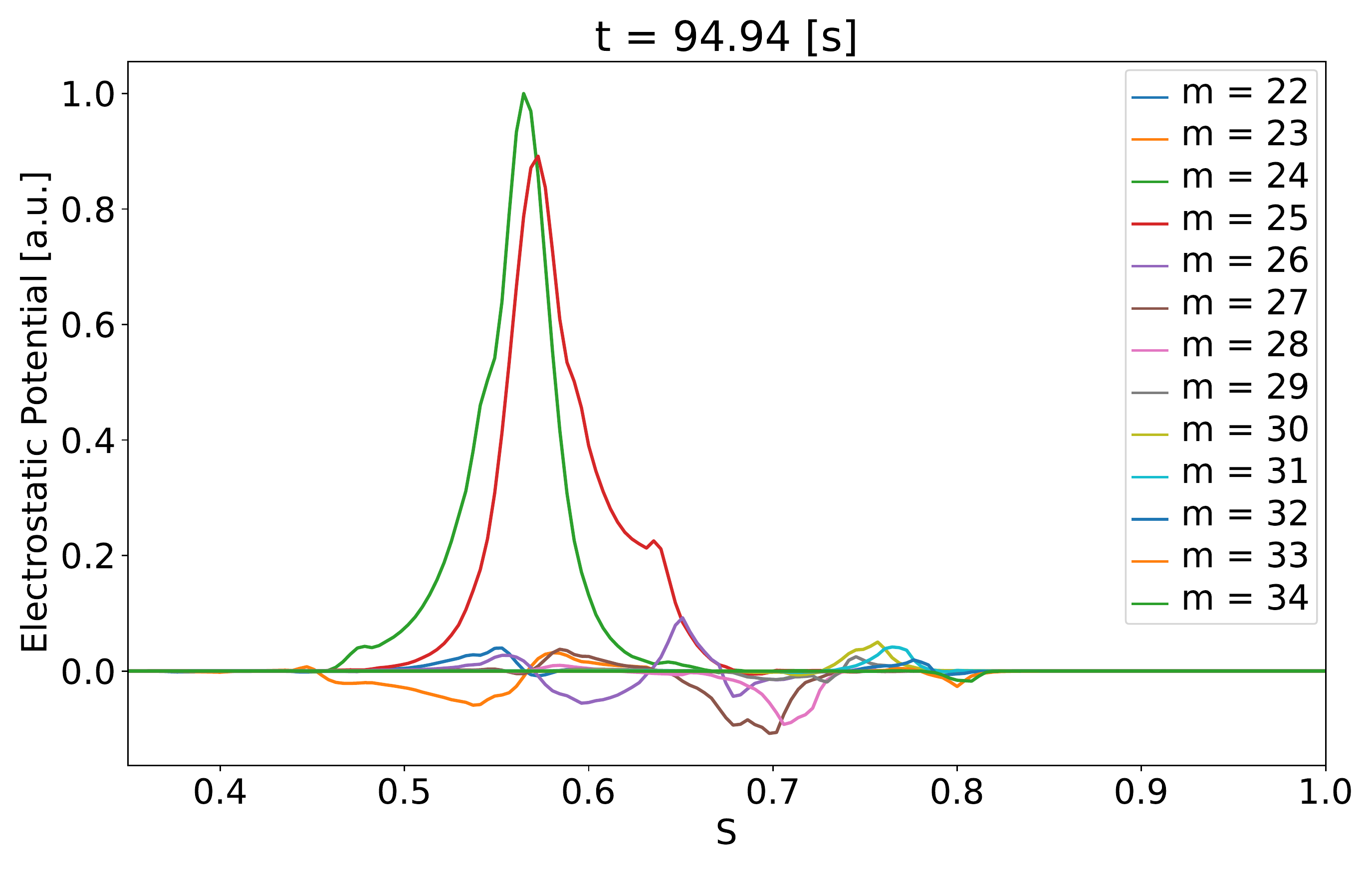}
  \caption{Mode structure for $n=25$, $m=(24,25)$ mode for $t=94.94$s}
  \label{fig:JINTRAC_EFS}
\end{figure}

\begin{figure}[ht]
  \centering
  \includegraphics[width=0.5\textwidth]{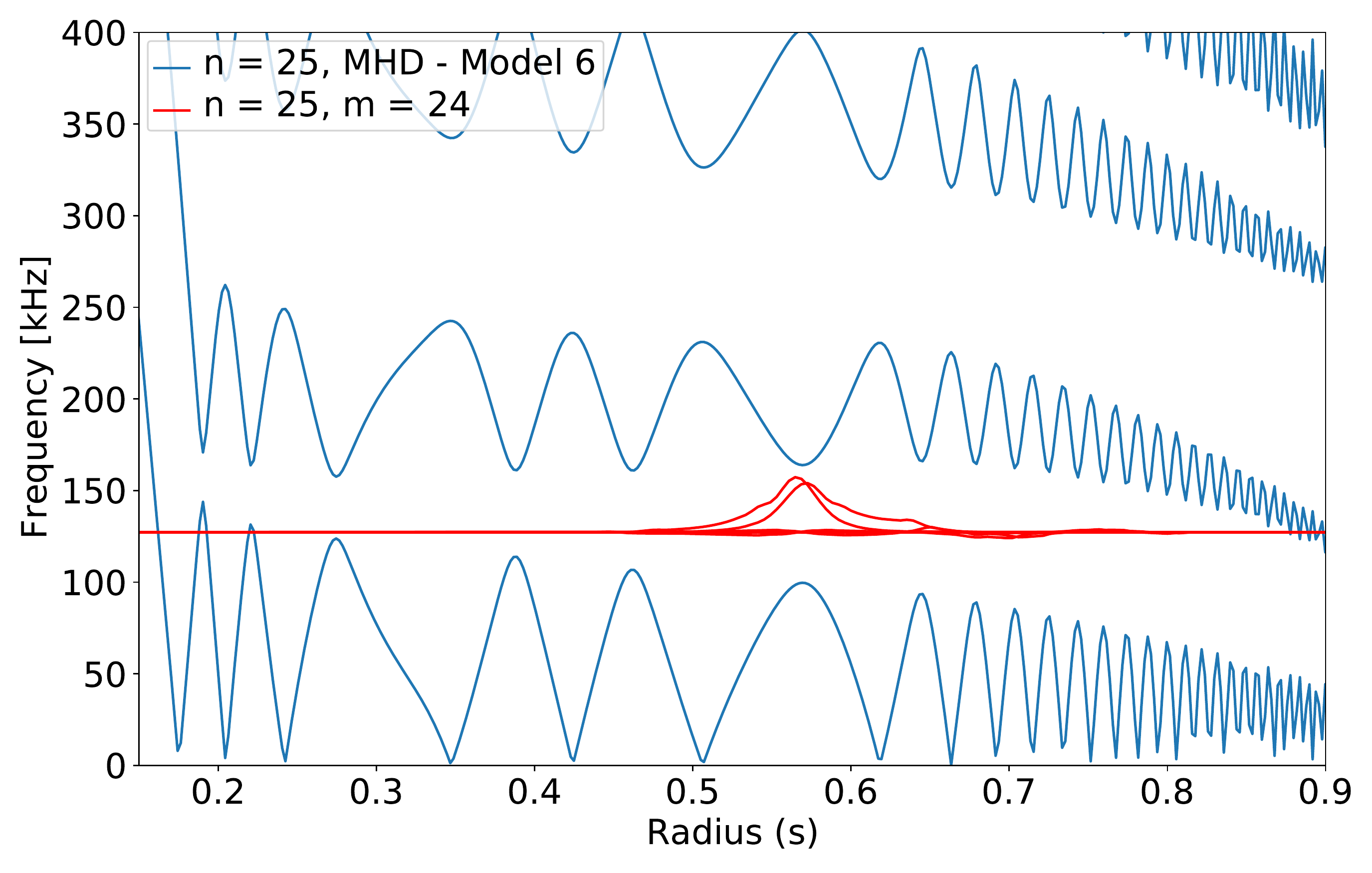}
  \caption{SAW continuum with the global TAE $n=25$, $m=(24,25)$ mode structure}
  \label{fig:JINTRAC_CONTINUUM_EFS}
\end{figure}

For completeness the mode structure is presented in Fig.\refeq{fig:JINTRAC_EFS}. One can see that despite the high toroidal mode number, the mode is quite global due to the flat shear around $s=0.5$. However, as Fig.\refeq{fig:JINTRAC_CONTINUUM_EFS} shows, the continuum damping is negligible since the mode touches the continuum at s=0.9 where all poloidal harmonics with significant relative amplitude have decayed. Thus, just other non-local damping effects (Ion/electron/radiative) are contributing via the wide mode structure to the overall mode damping.

\subsection{METIS - 130012/02}
Using the METIS scenario, the workflow demonstrates all the capabilities of a complete and automatic tool for linear stability inside a tokamak. Starting from an overview of the TAE locations and frequencies (via model 5) to a more in depth analysis via the local (model 4) and then global solver (model 1/2) was performed. First alpha particles are excluded to determine the damping rates, and then the effects of alpha particles are added. Also, a convergence test was performed in order to investigate and optimize the computational resources for the various analysis steps.

\begin{figure}[ht]
  \centering
  \includegraphics[width=0.5\textwidth]{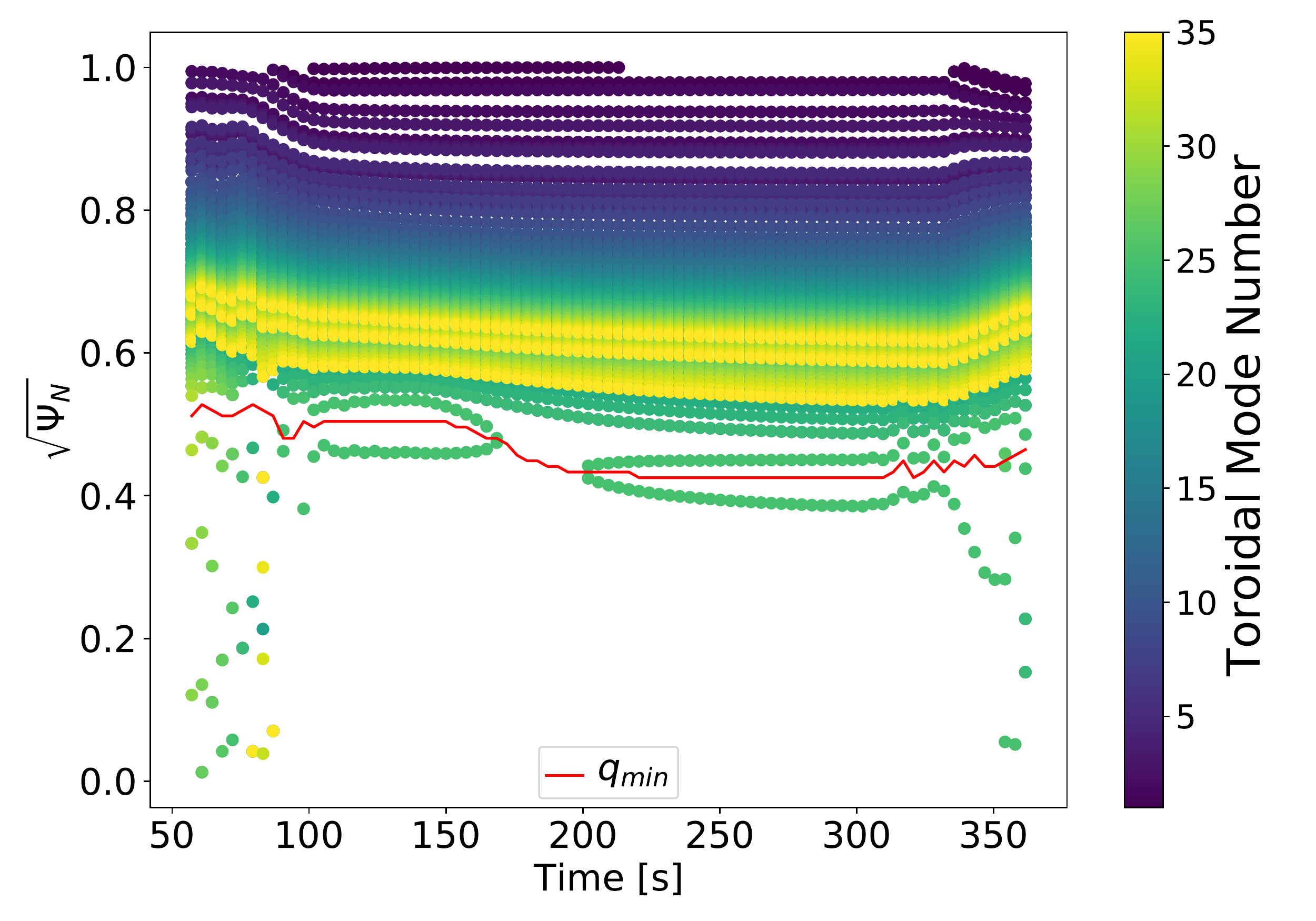}
  \caption{Radial location of modes with toroidal mode number ($n$) between 1 and 35 tracked over the entire scenario.}
  \label{fig:radial_location_metis}
\end{figure}

Due to the ability of model 5  to perform basic analytical AE calculations in a short amount of time ($<$1s - mode - time step), a scan over the whole METIS scenario (106 time points) was performed taking $n=1$ \dots $35$ and $m=n+1$ \dots $m=n+5$ for a total of $12014$ modes, averaging about $137$ modes per time point. In Fig.\refeq{fig:radial_location_metis} a plot of all these modes positions over time is given. Pointed in the graph is also the $q\approx 1$ value for each time point. The slight inversion of the q-profile around $\sqrt{\psi_N} \approx 0.5$ in the flat top part of the scenario ($> 90$s) obviously leads to two different TAE branches with the same mode numbers. Also, after  ($> 90$s)  no TAEs are found in the core ($s<0.4$). This is a consequence of the extremely small magnetic shear in the core assumed by the METIS-given equilibrium.

\begin{figure}[ht]
  \centering
  \includegraphics[width=0.5\textwidth]{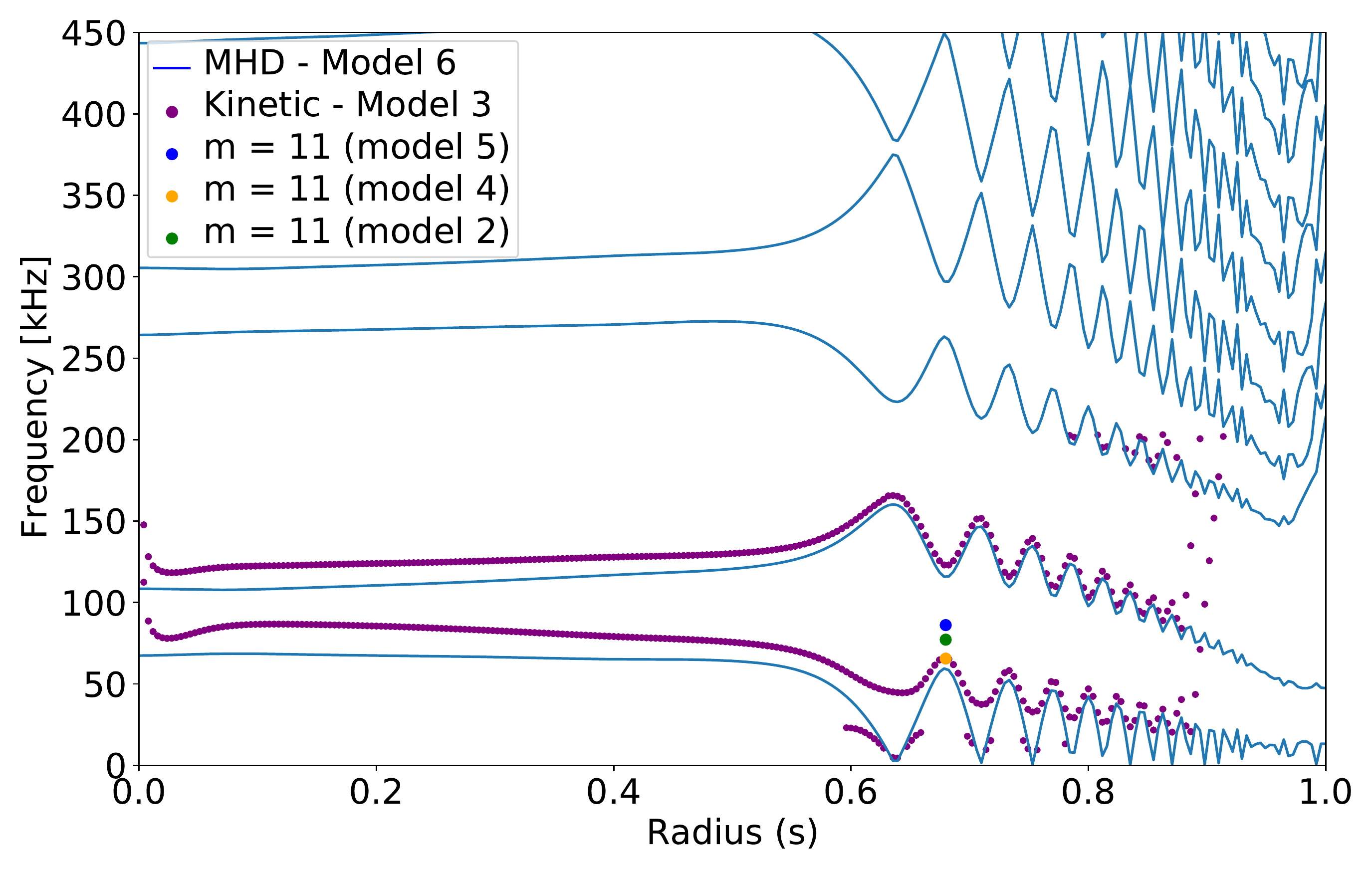}
  \caption{Continuum for $n=10$, $m=11$, with blue lines the MHD (Model 6) and with orange dots Kinetic (Model 3) at $t=90$s. The dots indicate the TAE frequencies as calculated by model 5 (analytical), 4 (local kinetic) and 2 (global kinetic).}
  \label{fig:continuum_metis}
\end{figure}

In Fig.\refeq{fig:continuum_metis} the reduced SAW continumm and the kinetic continuum for $n=10$  are  plotted. The results for one TAE ($m=11,12$) using the model hierarchy are added. As in the previous case, the different models give slightly different TAE frequencies (coloured dots in fig \refeq{fig:continuum_metis} ), in line with the discussion above. \\

Since the SAW continuum in Fig.\refeq{fig:continuum_metis} shows an almost open TAE gap, it is investigated how many poloidal  harmonics are needed for the convergence of this mid-n (n=10) TAE. This test is also of significant importance for solver validation inside LIGKA and the EP-WF. Several runs were done by looking at the same mode $(n,m)=(10,11)$ using a different number of poloidal harmonics.
\begin{figure}[ht]
  \centering
  \includegraphics[width=0.4\textwidth]{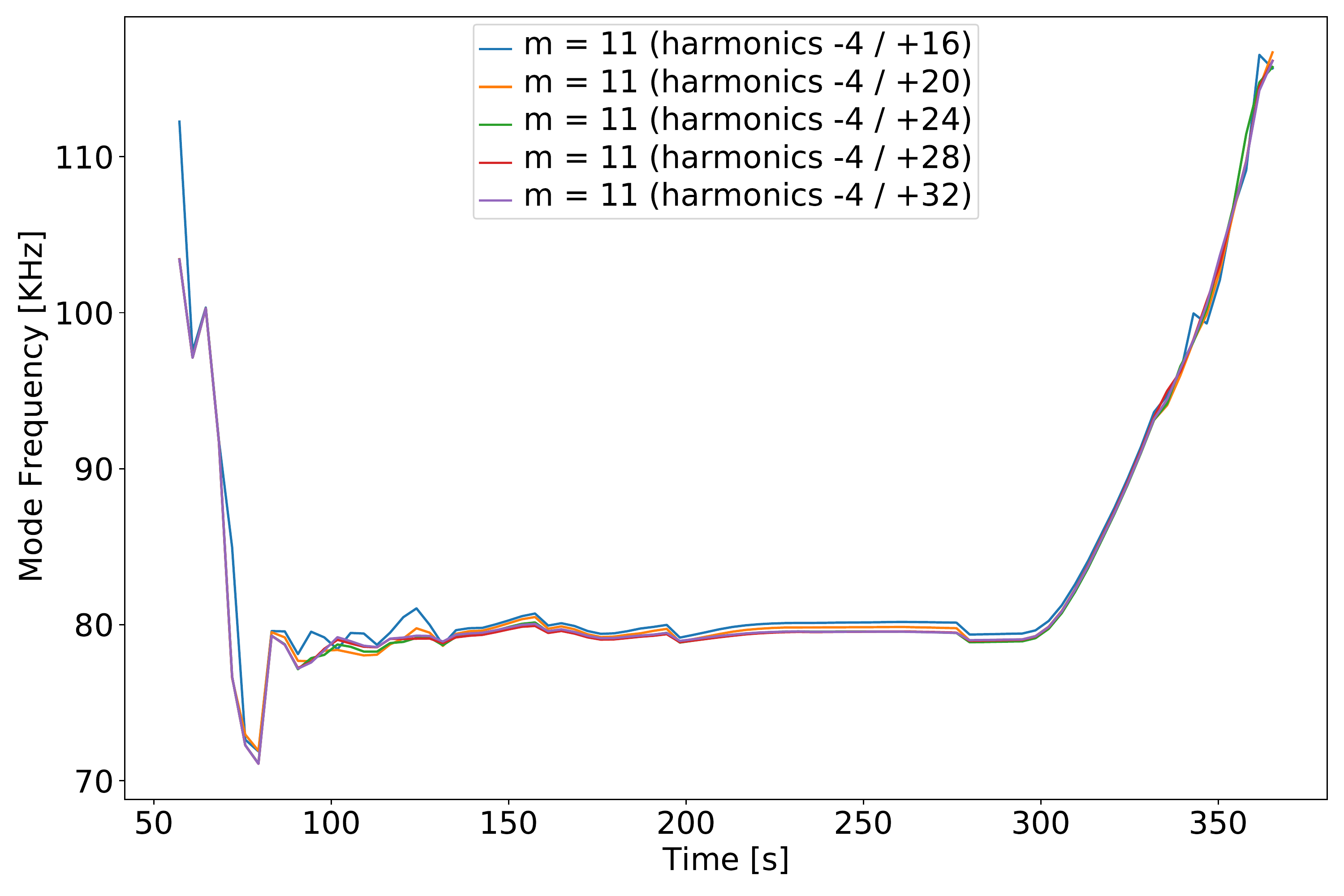}
  \caption{Frequency scan for $n=10$, $m=11$, for different numbers of poloidal harmonics tracked for the duration of the scenario. }
  \label{fig:frequency_harmonics}
\end{figure}
\begin{figure}[ht]
  \centering
  \includegraphics[width=0.4\textwidth]{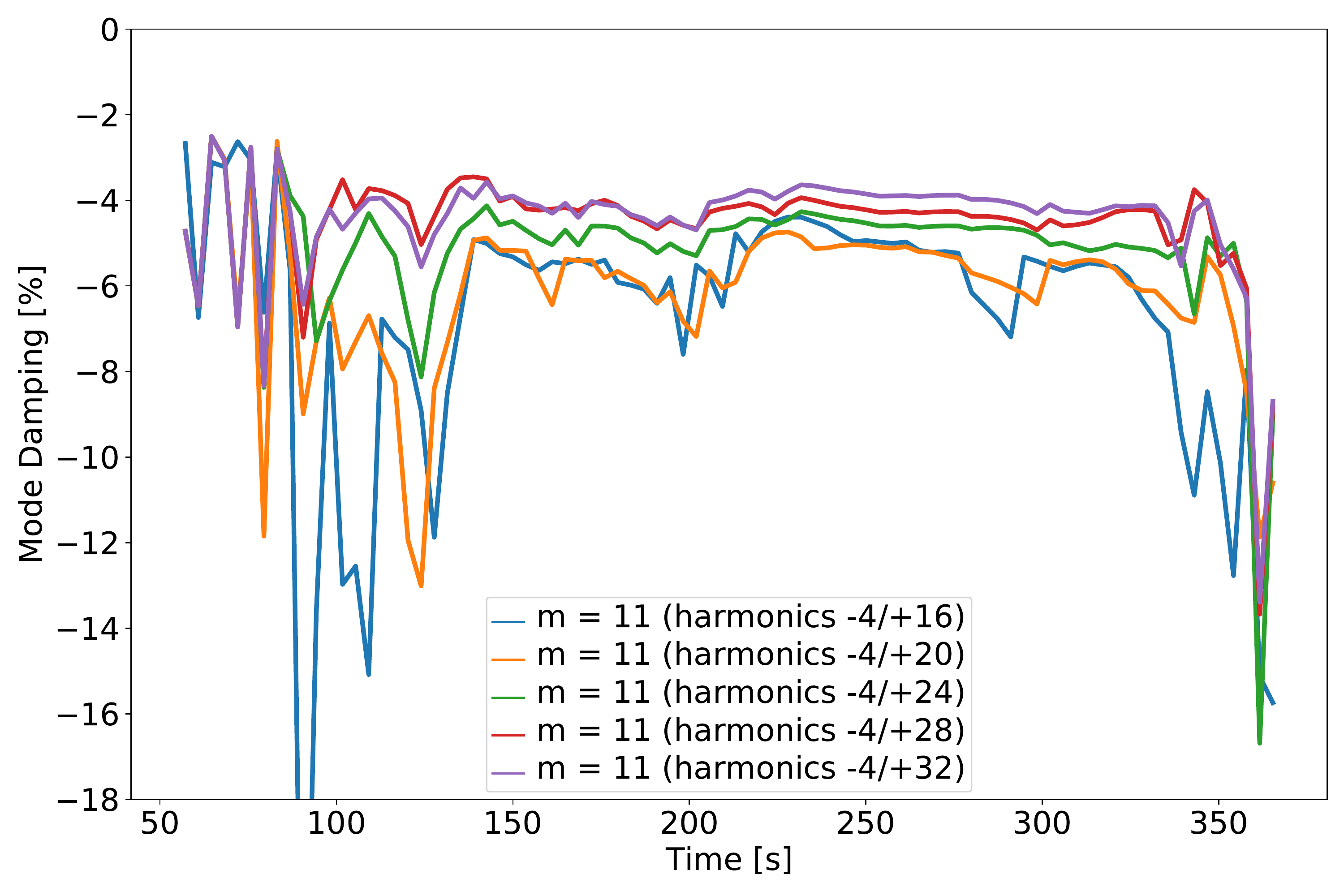}
  \caption{Growth rate for $n=10$, $m=11$, for different numbers of poloidal harmonics tracked throughout the whole scenario. }
  \label{fig:growth_harmonics}
\end{figure}
\begin{figure}[ht]
  \centering
  \includegraphics[width=0.4\textwidth]{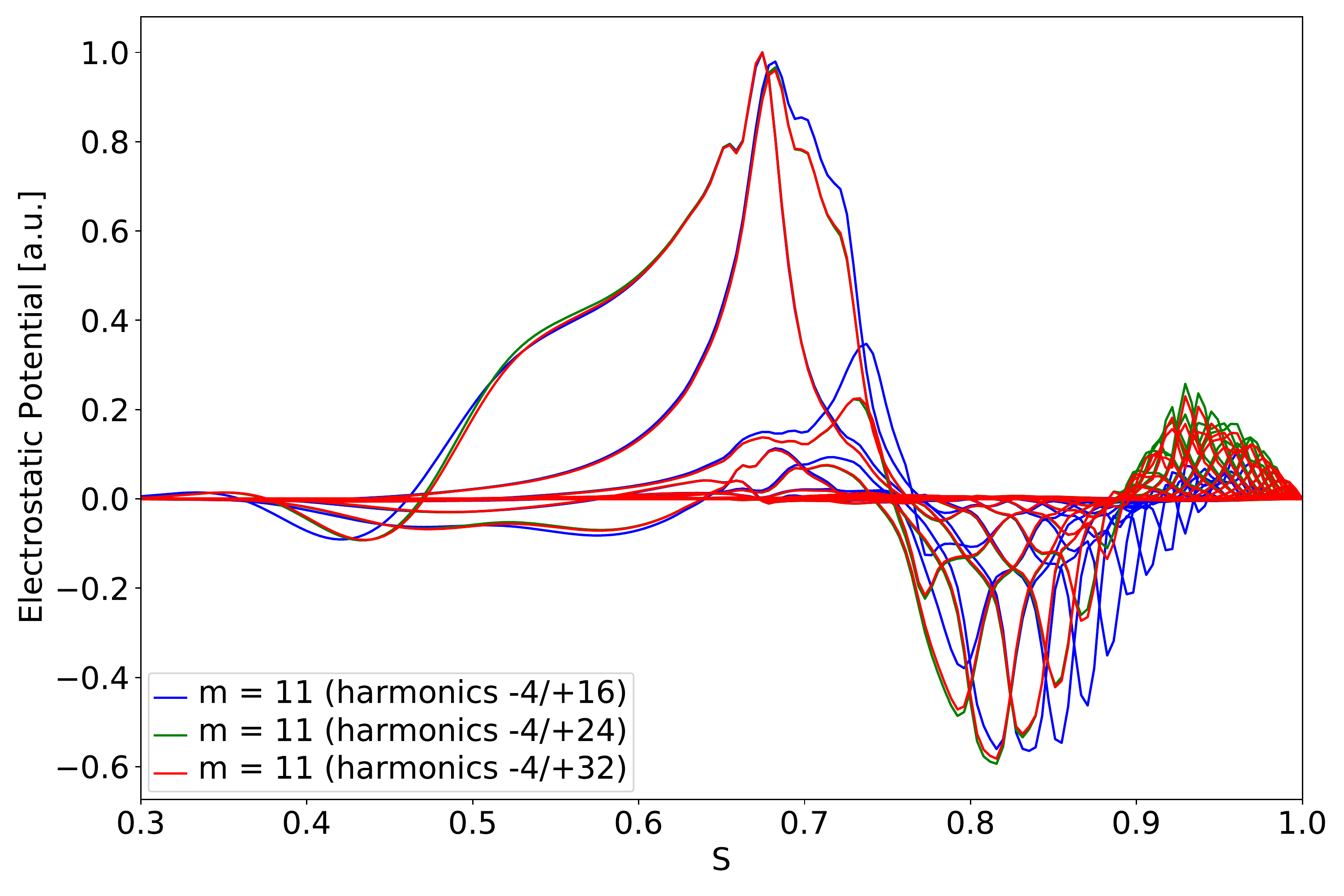}
  \caption{Mode structures of $n=10$, $m=11$, for different numbers of poloidal harmonics at $t=90$s. }
  \label{fig:mode_structure_harmonics}\end{figure}

In Fig.\refeq{fig:frequency_harmonics} the mode frequencies calculated using different numbers of poloidal harmonics are shown. The frequency differences between the different runs are not significant. During the ramp-up ($50<=t<=100$s) and ramp-down ($300<=t<=400$s) the variation in frequency is determined by the evolution of the density, temperature and safety profile. Note that the time step size for equilibria and profiles in this scenario is $\approx 3$s.  For flat-top phase of the scenario with stable profiles, almost no changes in the TAE frequency are observed.

In Fig.\refeq{fig:growth_harmonics} the damping rate of the same mode is is plotted, similarly to Fig.\refeq{fig:frequency_harmonics}. In this case, if the number of poloidal harmonics is varied, significant differences are observed. As it can be seen, the number of poloidal harmonics for convergence depends on the equilibrium, \ie{} particularly at the end of the ramp-up phase non-local continumm damping  towards the plasma edge  requires many poloidal harmonics to be included. This difference is particularly pronounced in the most unstable (profile wise) regions of the scenario: ramp-up and ramp-down. Depicted in the graph with blue we use 21 harmonics (from $m=n-4$ to $m=n+16$). One can see that as more harmonics are added, the results become more and more weakly damped. In the flat top region of the scenario, the difference is not that significant $\pm 0.5\%$, whereas in the ramp-up phase the damping stronlgy depends on the number of poloidal harmonics ( $\approx -25\%$). Adding more harmonics leads eventually to a converged damping rate, especially in the ramp-up phase (90-110s). Obviously, the mode number $n=10$ was chosen as the most challenging case for this scenario, since the global TAE touches the continumm close to the plasma edge, explaining the slow convergence behaviour: as the number of poloidal harmonics is increased, the interaction with the continuum is radially further and further separated from the main TAE location, deacreasing the damping rate (see fig.\refeq{fig:growth_harmonics}).
The continuum intersection moves to the edge and thus smaller and smaller poloidal harmonics interact with the (not yet resolved) continuum, until it is fully resolved ($-4 / +28$ and $-4 / +32$).\\

\begin{figure}[ht]
  \centering
  \includegraphics[width=0.5\textwidth]{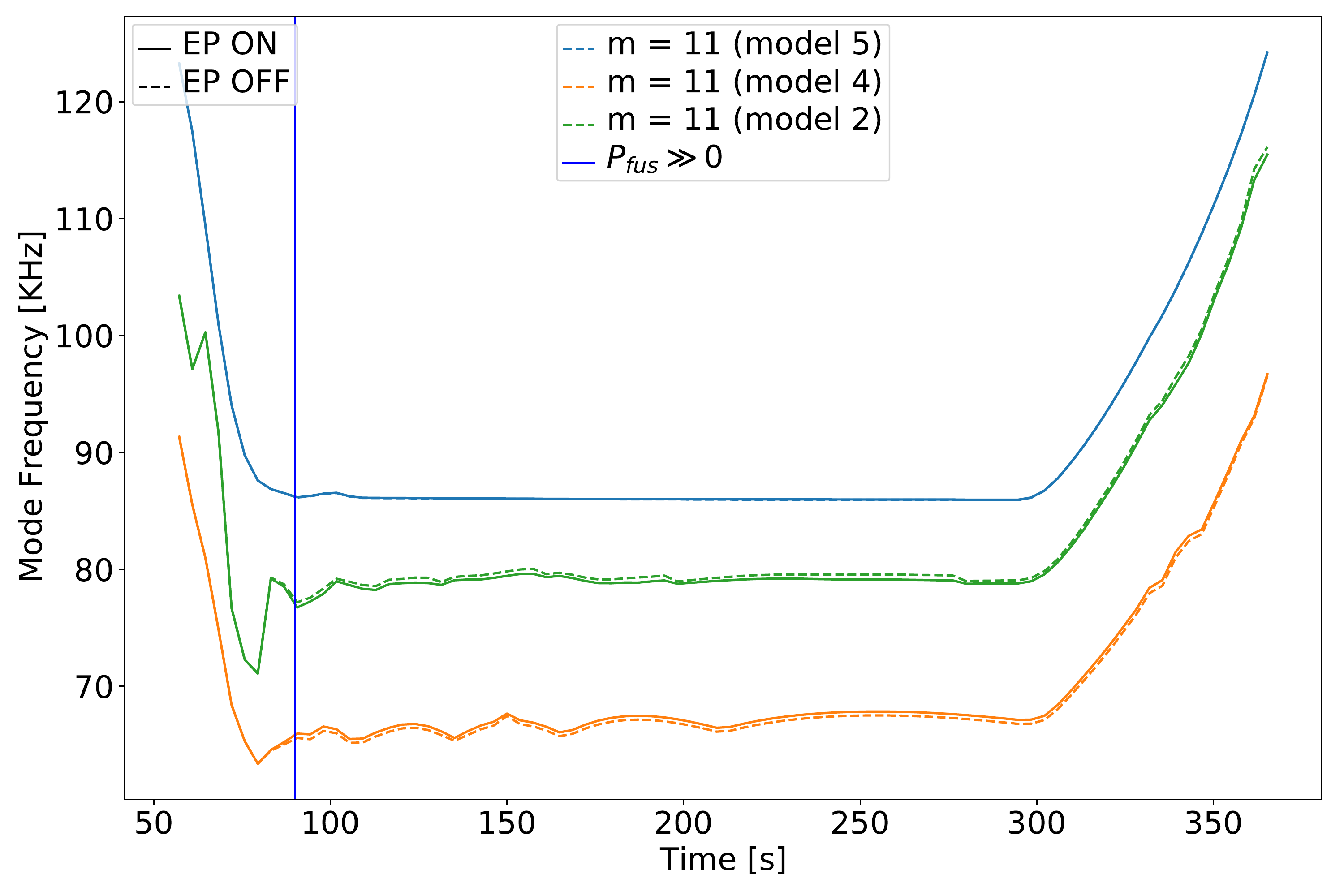}
  \caption{Frequency for the $n=10$, $m=11$ mode as function of time. Results with (solid) and without (dashed) alpha particles are shown.}
  \label{fig:freq_alphas_no_metis}
\end{figure}

\begin{figure}[ht]
  \centering
  \includegraphics[width=0.5\textwidth]{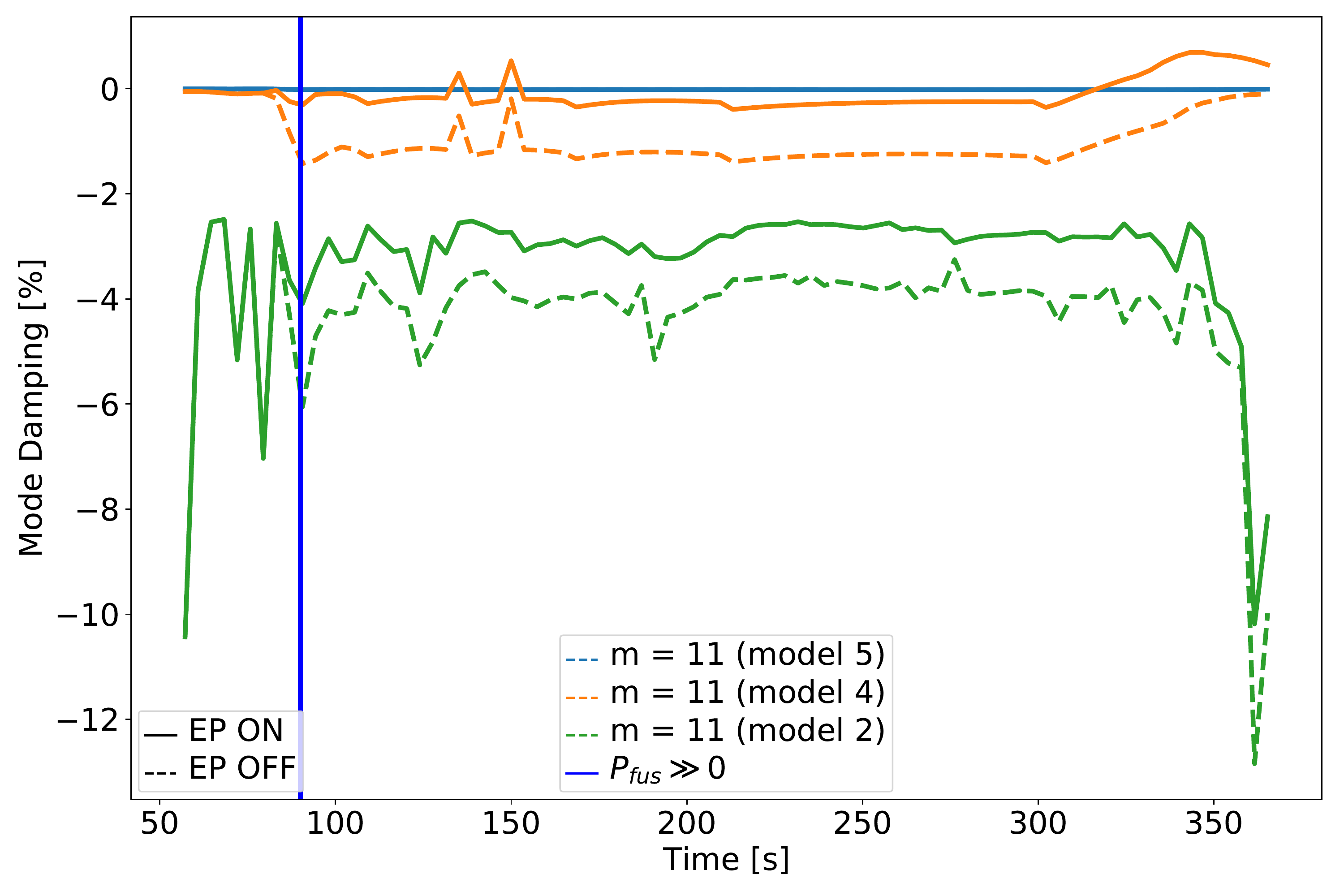}
  \caption{Damping rate for the $n=10$, $m=11$ mode as a function of time along the whole scenario. Results with (solid) and without (dashed) alpha particles are shown.}
  \label{fig:growth_alphas_no_metis}
\end{figure}

Next, alpha particles coming from the fusion reactions were added, as given by the scenario simulation. For simplicity, an effective hot Maxwellian was chosen, conistent with the EP pressure and density as given by the transport code.
In Figures \ref{fig:freq_alphas_no_metis} and \ref{fig:growth_alphas_no_metis}, the frequencies and damping rates for the same  $n=10$ and $(m=11,12)$ TAE for the whole time domain ($t=45 - 360$s) are given. Based on the previous analysis, the poloidal harmonics included in this simulation was $-4 / +32$ that was found to be the necessary range needed for convergence. Again, the various models are run and compared (dashed lines: no alphas, solid: alphas). The blue vertical line on the graphs indicate the moment when a significant population of alpha particles starts to be  present in the plasma due to fusion processes,

As expected, the frequency did not shift by a significant margin (Fig. \refeq{fig:freq_alphas_no_metis}) between runs with alphas or without.

Fig.\refeq{fig:growth_alphas_no_metis} shows the damping rate for both cases, with and without alpha particles taken into consideration. Using model 4 (local solver - orange line) one can see that there is a small difference between the cases with and without EPs. Moreover, this local model suggests that in the ramp down, the TAE becomes unstable in the presence of alphas. Using the global solver (model 1/2 - green line) one can see that this is not the case, due to the non-local damping mechanisms. At $t\approx 90$s  when alphas are turned on, the damping rates start to differ. However,  while the alphas do reduce damping, the TAE does not become  unstable throughout the whole scenario. The same is true for other n's (tested for $n=10$...$30$) showing that this extremely (unrealistically) flat q-profile results in a decoupling of TAE locations and steep EP gradient regions, leading to a stable TAE spectrum. Obviously small equilibrium deviations can change this, particularly assuming different models for density peaking. Using the established workflow sensitivity scans based on slightly differing assumptions in the scenario modelling process can be systematically investigated in the future.

  \section{Summary and outlook}
\label{sec:conclusions}

Our study aimed to develop an automated workflow for time-dependent EP transport analysis. The focus in this paper was on linear properties of TAEs in a torus, specifically in ITER predictive scenarios. A demonstration of the effectiveness of this workflow in identifying the key linear properties of modes such as damping rates, frequency, location and structure was performed.

The workflow together with the associated physics codes, have demonstrated the ability to generate hierarchical physics results. By utilizing efficient computational methods, reduced models and ensuring that all code and data are version-controlled (via IMAS) and documented we were able to produce phyically consistent and reproducible results. As an overview, in the table below, detailed information about the runtime of the test cases discussed above are presented.
\begin{table}[ht]
  \resizebox{0.5\textwidth}{!}{\begin{tabular}{|c|c|c|c|c|}
    \hline
    Case  & Run Scope               & Timeslices & Modes & Procs                 \\ \hline
    ASTRA & local + global scan     & 1          & 175   & local = 1, global = 8 \\ \hline
    ASTRA & local - continuum       & 1          & 3     & 1                     \\ \hline
    METIS & local scan              & 106        & 12014 & 1                     \\ \hline
    METIS & local + global (-4/+16) & 106        & 1     & local = 1, global = 8 \\ \hline
    METIS & local + global (-4/+20) & 106        & 1     & local = 1, global = 8 \\ \hline
    METIS & local + global (-4/+28) & 106        & 1     & local = 1, global = 8 \\ \hline
    METIS & local + global (-4/+32) & 106        & 1     & local = 1, global = 8 \\ \hline
    \end{tabular}}
    \newline
    \vspace*{0.5 cm}
    \newline
    \resizebox{0.5\textwidth}{!}{\begin{tabular}{|c|c|c|c|c|}
      \hline
      Case  & Run Scope               & Total Time & Time/timeslice & Time/mode     \\ \hline
      ASTRA & local + global scan     & 09:18:06   & 09:18:06       & 00:04:00      \\ \hline
      ASTRA & local - continuum       & 00:25:04   & 00:25:24       & 00:08:33      \\ \hline
      METIS & local scan              & 01:00:24   & 00:00:34       & \textless{}1s \\ \hline
      METIS & local + global (-4/+16) & 04:36:41   & 00:02:36       & -             \\ \hline
      METIS & local + global (-4/+20) & 06:20:02   & 00:03:35       & -             \\ \hline
      METIS & local + global (-4/+28) & 10:31:06   & 00:05:56       & -             \\ \hline
      METIS & local + global (-4/+32) & 13:00:35   & 00:07:23       & -             \\ \hline
      \end{tabular}}
  \caption{Run times for different cases presented in this paper. The computational resources are also shown in terms of processes being used for each run. In the total run time for the equilibrium code at each timeslice is also included.}
  \end{table}

The successful application of this approach represents a fundamental and challenging first step towards developing more sophisticated tools for analyzing EP transport. These type of tools will be critical for optimizing the performance of fusion reactors, such as ITER. This work will contribute directly to various reduced EP transport models.

  \section*{Acknowledgements}
This work has been carried out within the framework of the EUROfusion Consortium, funded by the European Union via the Euratom Research and Training Programme (Grant Agreement No. 101052200- EUROfusion), partially within the framework of the Enabling research project ’ATEP’. The views and opinions expressed herein do not necessarily reflect those of the European Commission.

ITER is the Nuclear Facility INB no. 174. The views and opinions expressed herein do not necessarily reflect those of the ITER Organisation. Part of this work has been performed under ITER IO Contract No. IO/18/CT/ 4300001657.
  \section*{References}
  \bibliographystyle{unsrt}
  \bibliography{article_alin}
\end{document}